\documentclass[fleqn,usenatbib]{mnras}

\usepackage{natbib, aas_macros}
\usepackage{ulem}
\usepackage[dvips]{graphicx}
\usepackage{wrapfig}
\graphicspath{{./figs/}}
\usepackage{color}
\usepackage{amsmath}
\usepackage{amssymb}

\newcommand{\Msun}{{\rm  M_{\odot}}}

\newcommand{\Zsun}{Z_{\odot}}

\newcommand{\Mh}{M_{\rm h}}

\newcommand{\Msunyr}{{\rm \Msun~{\rm yr^{-1}}}}

\newcommand{\Mstar}{{\rm M_{star}}}
\newcommand{\Muv}{{M_{\rm UV}}}

\title[Forever22: the first bright galaxies] 
{FOREVER22: the first bright galaxies with population III stars at redshifts $z \simeq 10-20$ and comparisons with JWST data}
\author[Yajima et al. ]{Hidenobu Yajima$^{1}$\thanks{E-mail: yajima@ccs.tsukuba.ac.jp}, Makito Abe$^{1}$, Hajime Fukushima$^{1}$, Yoshiaki Ono$^{2}$, Yuichi Harikane$^{2}$,  
\newauthor Masami Ouchi$^{2,3,4}$, Takuya Hashimoto$^{5}$, Sadegh Khochfar$^{6}$\\
$^{1}$Center for Computational Sciences, University of Tsukuba, Ten-nodai, 1-1-1 Tsukuba, Ibaraki 305-8577, Japan\\
$^{2}$Institute for Cosmic Ray Research, The University of Tokyo, 5-1-5 Kashiwanoha, Kashiwa, Chiba 277-8582, Japan\\
$^{3}$ Kavli IPMU (WPI), The University of Tokyo, 5-1-5 Kashiwanoha, Kashiwa, Chiba, 277-8583, Japan \\
$^{4}$National Astronomical Observatory of Japan, 2-21-1 Osawa, Mitaka, Tokyo 181-8588, Japan\\
$^{5}$ Tomonaga Center for the History of the Universe (TCHoU), Faculty of Pure and Applied Sciences, University of Tsukuba, \\
Tsukuba, Ibaraki 305-8571, Japan\\
$^{6}$Institute for Astronomy, University of Edinburgh, Royal Observatory, Edinburgh, EH9 3HJ, UK\\
}

\begin{document}

\date{Accepted ?; Received ??; in original form ???}

\pagerange{\pageref{firstpage}--\pageref{lastpage}} \pubyear{2008}

\maketitle

\label{firstpage}

%
%
\begin{abstract}
We study the formation of the first galaxies in overdense regions modelled by the FORmation and EVolution of galaxies in Extremely overdense Regions motivated by SSA22 (FOREVER22) simulation project. 
Our simulations successfully reproduce the star formation rates and the $M_{\rm UV}-M_{\rm star}$ relations of candidate galaxies at $z \sim 10-14$ observed by the James Webb Space Telescope (JWST). We suggest that the observed galaxies are hosted by dark-matter haloes with $M_{\rm 
 h} \gtrsim 10^{10}~\Msun$ and are in short-period starburst phases. On the other hand, even simulated massive galaxies in overdense regions cannot reproduce the intense star formation rates and the large stellar masses of observed candidates at $z \sim 16$. 
 Also, we show that the contribution of population III stars to the UV flux decreases as the stellar mass increases and it is a few percent for galaxies with $M_{\rm star} \sim 10^{7}~\Msun$. Therefore, a part of the observed flux by JWST could be the light from population III stars.
 Our simulations suggest that the UV flux can be dominated by population III stars and the UV-slope shows $\beta \lesssim -3$ if future observations would reach galaxies with $M_{\rm stars} \sim 10^{5}~\Msun$ at $z \sim 20$ of which the mass fraction of population III stars can be greater than 10 percent. 
\end{abstract}
%
%

%
%
\section{Introduction}
Understanding galaxy formation is one of the central issues in current astrophysics. 
In particular, the first galaxies at redshifts beyond $z=10$ are the most likely drivers 
of cosmic reionization 
\citep{Yajima09, Yajima11, Yajima14c, Paardekooper13, Paardekooper15, Wise14, Arata19, Ma20, Rosdahl22} and hosts of the first massive black holes \citep{Regan09, Agarwal14, Yajima16a, Wise19, Latif22b}. Thus, revealing the formation of the first galaxies is of great importance.  
Using Lyman-$\alpha$ lines, a lot of galaxies at $z \lesssim 9$ have been identified  
\citep[e.g.,][]{Shibuya12, Ono12, Finkelstein13, Song16b, Ouchi18}.
However, the transmission of the Lyman-$\alpha$ line is reduced significantly as the neutral degree of the inter-galactic medium increases, resulting in the difficulty of galaxy observation beyond $z \sim 10$ \citep{Yajima18}. Recent submillimeter observations have successfully detected high-redshift galaxies at $z \lesssim 9$ via the detections of $\rm [C_{II}]~158~\mu m$ and $\rm [O_{III}]~88~\mu m$ lines \citep[e.g.,][]{Capak15, Inoue16, Hashimoto18, Tamura19}. The metal line observation is expected to be difficult if target galaxies exceed $z \sim 10$ because of insufficient metal enrichment \citep[e.g.,][]{Yoon22, Bakx22, Popping22}. 
Therefore, galaxies at $z \gtrsim 10$ have been investigated with Lyman-break technique \citep[e.g,][]{Oesch13, Oesch16, Bouwens19}. However, the number of samples has been limited and the spectroscopic confirmations have been difficult for the sensitivities of the telescopes with a reasonable integration time. 

These situations are drastically changing with observations by JWST. 
Using the data of the first cycle observation by JWST, high-redshift galaxies have been identified. \citet{Donnan23}  found 44 new candidate galaxies and estimated the UV luminosity functions at the redshifts $z=8-15$. \citet{Harikane23} found candidate galaxies at $z \sim 16$ with large stellar masses and star formation rates \citep[see also,][]{Naidu22}.
\citet{Furtak23} indicated that the candidate galaxies at $z \sim 9-16$ had properties of young galaxies with  ages $\sim 10-100~\rm Myr$ and very blue UV slope down to $\beta \sim -3$ \citep[see also,][]{Topping22, Cullen22}. 

 The properties of high-redshift galaxies at $z > 6$ have been investigated in various simulation projects as 
{\sc CoDa} \citep{Ocvirk16}, 
{\sc flares} \citep{Lovell21}, {\sc thesan} \citep{Kannan22a},  {\sc MilleniumTNG} \citep{Pakmor22}, {\sc universemachine} \citep{Behroozi19} and Santa Cruz model \citep{Gabrielpillai22}.
These simulations successfully reproduced statistical properties like luminosity functions of observed galaxies at $z \lesssim 8$. Also, some previous works provided theoretical predictions of galaxy properties at $z \gtrsim 10$ from the simulation results \citep[e.g.,][]{Behroozi20, Kannan22b}. While these simulations allow us to study the statistical natures of high-redshift galaxies with large cosmic volumes, it is still difficult to study evolution from mini-haloes hosting the population (Pop) III stars to massive galaxies due to the limited resolution. 

In previous theoretical studies, galaxy formation at $z \gtrsim 10$  proceeds with the formation of Pop III stars, the radiative feedback, and the metal enrichment via the first supernovae \citep[e.g.,][]{Maio11, Wise12a, Johnson13, Smith15, Xu16, Chiaki19}.
Due to the metal enrichment from Pop III stars, formation sites of new Pop III stars move from higher to lower density regions in large-scale structure \citep[e.g.,][]{Tornatore07, Pallottini14, Xu16, Liu20}.
Such numerical simulations bridging from population III stars in mini-haloes to first galaxies are still challenging. \citet{Jeon19} investigated the formation of first galaxies with the halo mass of $\Mh \sim 10^{9}~\Msun$ at $z=9$ and showed their observational properties. \citet{Abe21} studied the impact of the initial mass function of population III stars on the physical properties of first galaxies with $\Mh \sim 10^{8-9}~\Msun$. They showed that  inducing frequent pair-instability supernovae suppress the gas mass fraction and the star formation rates (SFRs) of the first galaxies significantly for the top-heavy initial mass function. The simulated halo masses in previous works have been limited to $\sim 10^{9}~\Msun$. Therefore, the emergent UV fluxes were too faint for the sensitivities of current telescopes. 

Considering the brightness of observed candidates at $z \gtrsim 10$, they can be hosted in massive haloes which likely form in overdense regions. 
In this work, we investigate galaxy formation in overdense regions in which the halo mass exceeds $10^{11}~\Msun$ at $z \sim 10$. To study the transition from population III to II stars, our simulations resolve mini-haloes and follow their growth up to the massive haloes.  

Our paper is organized as follows. 
Section 2 shows our methodology and the information about the simulation setup. In section 3, we show the star formation histories and compare them with the observational data by JWST. Also, we study the mass fraction of Pop III stars with regard to the total stellar mass. 
Finally, we summarize our results and discuss the 
limitations of our study in section 4. 

%
%
\begin{figure*}
\centering	\includegraphics[width=13cm]{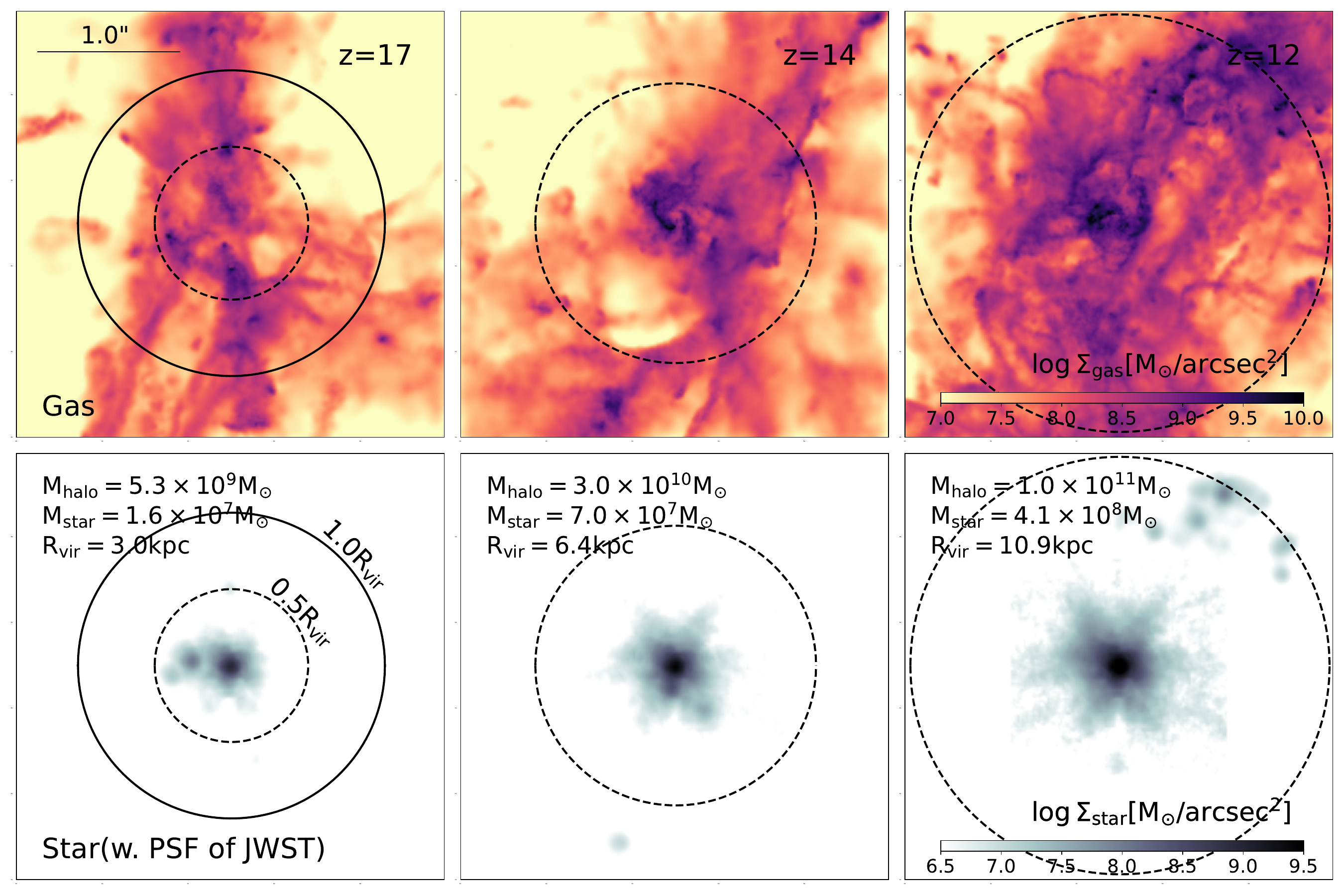}
	\caption{		
	Column density maps of gas and stars in the most massive progenitors at $z=12,14$ and $17$. The stellar distributions are smoothed with the point spread function of JWST. 
  Halo (stellar) masses are $5.3\times 10^{9}~(1.6\times 10^{7})~\Msun$ at $z=17$, $3.0\times 10^{10}~ (7.0 \times 10^{7})~\Msun$ at $z=14$, and $1.0 \times 10^{11}~(4.1 \times 10^{8})~\Msun$ at $z=12$.
 Solid and dashed circles represent virial radii and half virial radii, respectively. The size of the boxes is 3 arcsec, corresponding to $L=11.0, 9.8$ and $8.4$ physical kpc at $z=12, 14$ and $17$.
		 }
	\label{fig:2dmap}
\end{figure*}

\section{Methodology}

We use the results of our simulation project {\sc forever22} \citep{Yajima22} which focuses on protocluster regions in the cosmic volume of $(714~\rm cMpc)^{3}$. 
In this project, we use the {\sc gadget-3} code \citep{Springel05e} with sub-grid models developed in the {\sc owls} project \citep{Schaye10} and the {\sc fiby} project \citep{Johnson13}. Besides, we newly updated the code by adding the photo-ionization feedback, the radiation pressure on dust, dust growth/destruction, black hole growth, and its feedback \citep[see more,][]{Yajima22}.
The project consists of zoom-in simulations with three different levels of the mass resolution and the size of zoom-in regions: PCR (Proto-Cluster Region;  $V= (28.6~{\rm cMpc})^{3}$, SPH particle mass, $m_{\rm{SPH}} = 4.1 \times 10^{6}~\Msun$ and final redshift, $z_{\rm end}=2.0$), BCG (Brightest proto-Cluster Galaxy; $V \sim (10~{\rm cMpc})^{3} $, $m_{\rm SPH} = 5.0\times10^{5}~\Msun$ and $z_{\rm end}=4.0$ ), and First ( $V \sim (3~{\rm cMpc})^{3} $, $m_{\rm SPH} = 7.9 \times 10^{3}~\Msun$ and $z_{\rm end}=9.5$). The PCR runs reproduce the observed star formation rate densities of protoclusters at $z \sim 2-6$. Also, we confirmed that the mean density fields reproduced the observed stellar mass functions, main sequences of star formation, gas fractions, and metallicities of galaxies as a function of stellar mass well \citep{Yajima22}.  
In this work, we use First runs (First0 and First1 runs) in which the most massive halo reaches $\Mh=4.8 \times 10^{11}~ \Msun$ at $z=9.5$.
The cosmological parameters are still under debate \citep{Komatsu11, Planck20, Freedman21}.
Considering the changing history of the parameter and Hubble parameter tension \citep{Freedman21}, we adopt the cosmological parameters as $\Omega_{\rm M}=0.3, \Omega_{\rm b}=0.045, \Omega_{\rm \Lambda}=0.7,\sigma_{\rm 8}=0.82$ and $h=0.7$.

In this work, we consider both Pop II and III stars. If the metallicity is lower than a critical value, the initial mass function (IMF) is likely to be a top-heavy \citep[e.g,][]{Chon21}. Besides, the effective temperature of Pop III stars is high $T \sim 10^{5}~\rm K$ \citep{Schaerer02}. Therefore, Pop III stars can be strong sources of radiative and supernova (SN) feedback. 
We set the critical gas metallicity $Z=1.5 \times 10^{-4}~\Zsun$ below which Pop III stars form \citep{Omukai05, Frebel07, Chon21}. Although the critical metallicity is still under debate, \citet{Abe21} suggested that the physical properties of the first galaxies did not depend on it sensitively \citep[see also,][]{Maio10}.   
We assume that the IMF of Pop III stars is $dn \propto M^{-2.35}dM$ with the mass range $21-500~\Msun$ while that of Pop II is Chabrier IMF with the range $0.1-100~\Msun$. 
Because of the expensive calculation costs for the first-star formation with radiative and magnetic feedback, the IMF of Pop III stars is still under debate \citep{Stacy14, Susa14, Hirano15, Sugimura20, Wollenberg20, Latif22}. Therefore, we adopt a simple power-law function for the IMF of Pop III stars. 

In evaluating the star formation rate, we consider the star formation model based on the observed Kennicutt-Schmidt law which was developed in \citet{Schaye08}. The local SFR is measured as $\dot{m}_{\rm star} \propto m_{\rm gas}A \left( 1~{\rm \Msun~pc^{-2}}\right)^{-n}\left( \frac{\gamma}{G}f_{\rm g} P\right)^{(n-1)/2}$, where $m_{\rm gas}$ is the mass of a gas particle, $\gamma=5/3$ is the ratio of specific heats, $f_{\rm g}$ is the gas mass fraction in the galactic disc, and $P$ is the total ISM pressure. Here, we set $A=1.5 \times 10^{-4}~\Msunyr~\rm kpc^{-2}$ and $\gamma=1.4$ for $n_{\rm H} < 10^{3}~\rm cm^{-3}$ and $\gamma=2.0$ for $n_{\rm H} \ge 10^{3}~\rm cm^{-3}$. The star formation model is the same as in {\sc eagle} simulation project \citep{Schaye15}. Star formation occurs if local gas density exceeds $n_{\rm H}=n_{0}~{\rm cm^{-3}} \left( \frac{Z}{0.002}\right)^{-0.64}$, where we set $n_{0}=10.0$ for First runs. In the estimation of the net cooling rate, we follow the non-equilibrium chemistry of primordial gas and the equilibrium state of metals from pre-calculated tables with {\sc cloudy} v07.02 code \citep{Ferland00}. 

Once massive stars form, they give UV radiation feedback to surrounding gas within their lifetime $\sim 10^{7}~\rm yr$. We take into account the photo-ionization process of hydrogen and the dissociation of hydrogen molecules. 
 We estimate the volume of the ionized region by taking the balance between the photon production rate and the total recombination rate as \citep[see the detail,][]{Abe21, Yajima22}  
 \begin{equation}
 \dot{N}_{\rm ion} = \sum_{i=1}^{n} \alpha_{\rm B} n_{\rm HII}^{i} n_{\rm e}^{i} \frac{m_{\rm gas}^{i}}{\rho_{\rm gas}^{i}},
 \end{equation}
 where $\dot{N}_{\rm ion}$ is the photon production rate of a stellar particle, $\alpha_{\rm B}$ is the case-B recombination coefficient, 
 $n_{\rm HII}^{i}$ and $n_{\rm e}^{i}$ are the ionized hydrogen and electron number densities of $i$-th SPH particle. 
In the ionized regions, the gas temperature is heated up to $3 \times 10^{4}~\rm K$, and star formation is prohibited. The dissociation rate of hydrogen molecules is evaluated based on the contributions of stars in the calculation box. First, we measure UV fluxes from stars with  distances to a target gas particle as 
\begin{equation}
J_{\rm LW, 21} = \sum_{i=1}^{n} f_{\rm LW} \left( \frac{r_{\rm i}}{\rm 1~kpc}\right)^{-2} \left(  \frac{m_{\rm *, i}}{10^{3}~\Msun} \right),
\end{equation}
where $J_{\rm LW, 21}$ is described in unit of $10^{-21}~\rm erg \; s^{-1} \; cm^{-2} \; Hz^{-1} \; str^{-1}$,
$r_{\rm i}$ is the distance from $i$-th stellar particle to a target gas particle and $m_{\rm *, i}$ is the mass of $i$-th stellar particle. 
Then, we take into account the self-shielding effect with the local $\rm H_{2}$ density and Jeans length \citep{Johnson13}:
\begin{equation}
N_{\rm H_{2}} = 2 \times 10^{15}~{\rm cm^{-2}}~ \left( \frac{f_{\rm H_{2}}}{10^{-6}}\right)
\left( \frac{n_{\rm H}}{10~\rm cm^{-3}} \right)^{1/2} \left( \frac{T}{10^{3}~\rm K} \right)^{1/2},
\end{equation}
where $f_{\rm H_{2}}$ is the fraction of $\rm H_{2}$, $n_{\rm H}$ is the hydrogen number density. 
We consider the shielding factor derived in \citet{Wolcott-Green11} as
\begin{equation}
\begin{split}
f_{\rm shield}(N_{\rm H_{2}}, T) = &\frac{0.965}{(1 + x/b_{5})^{1.1}}
+ \frac{0.035}{(1+ x)^{0.5}} \\
&~~~\times {\rm exp} \left[ -8.5 \times 10^{-4} (1+x)^{0.5} \right],
\end{split}
\end{equation}
where $x \equiv N_{\rm H_{2}}/5 \times 10^{14}~\rm cm^{-2}$ and $b_{5} \equiv b/10^{5}~\rm cm~s^{-1}$. 
Here $b$ is the Doppler broadening parameter, $b \equiv (k_{\rm B}T/m_{\rm H})^{1/2}$. 
Thus, we estimate the $\rm H_{2}$ dissociation rate ($\kappa_{\rm diss}$) by combining $J_{\rm LW, 21}$ and $f_{\rm shield}$ as $\kappa_{\rm diss} \propto f_{\rm shield} J_{\rm LW, 21}$. In addition, we consider the photodetachment process of $\rm H^{-}$ \citep{Shang10}. With the dissociation and formation rates, we evaluate $\rm H_{2}$ abundance and its radiative cooling rate which is a main factor in controlling the formation of Pop III stars in mini-haloes. 
 
When the age of a stellar particle reaches $10^{7}~\rm yr$,  supernova (SN) feedback turns on. Following the SN feedback model in \citet{DallaVecchia12}, we stochastically select a neighbouring gas particle and heat the temperature up to $10^{7.5}~\rm K$. This hot bubble rapidly expands and induces galactic wind, resulting in the suppression of star formation.

%
%

\section{Results}
 
\begin{figure}
\centering \includegraphics[width=8cm]{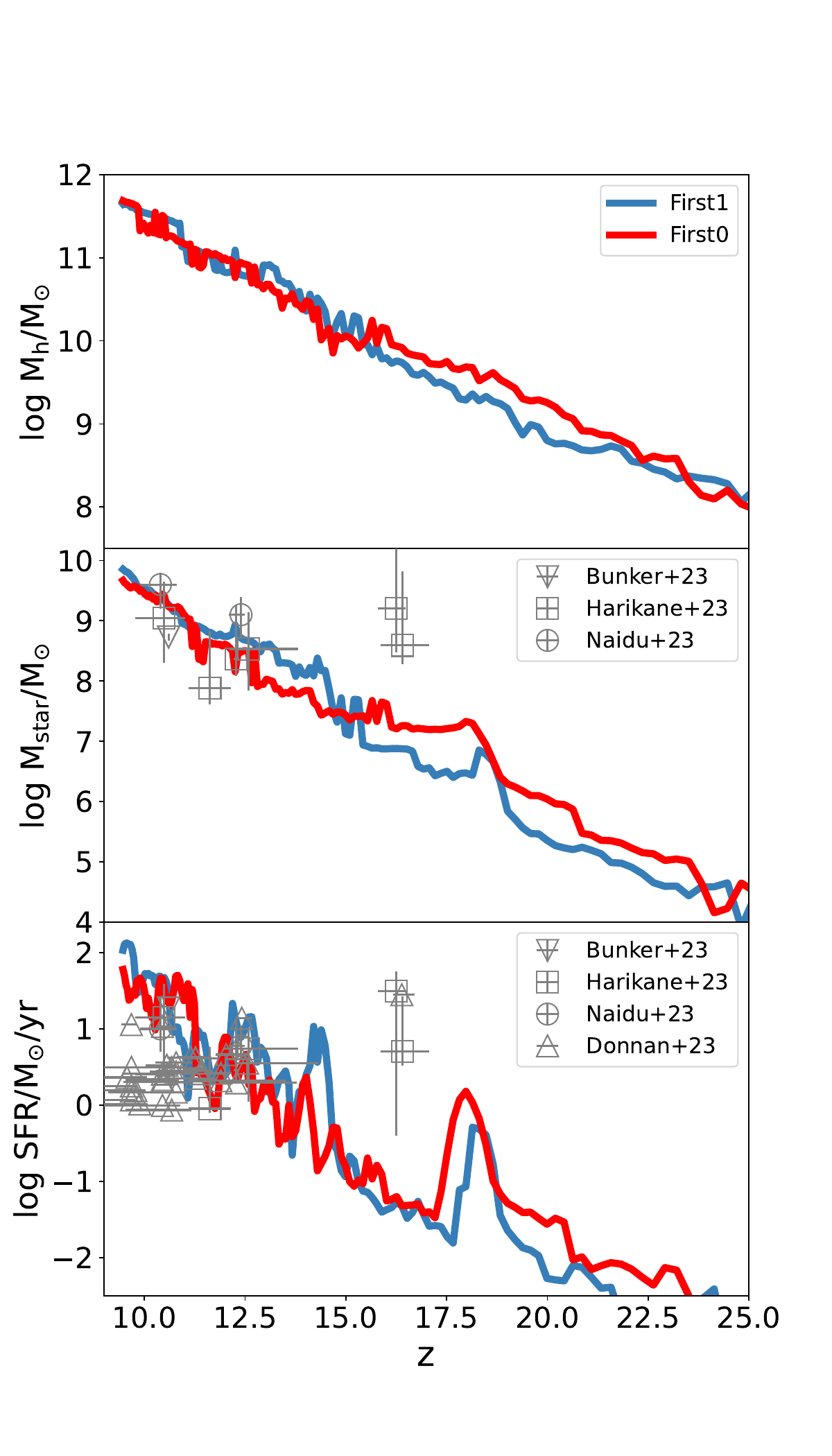}
	\caption{		
	Redshift evolution of the halo, stellar mass, and star formation rate. The red and blue lines represent the properties of the most massive progenitors in First0 and First1 runs, respectively. Open symbols show the observations: inverted triangles \citep{Jiang21}, squares \citep{Harikane23}, circles \citep{Naidu22}, and triangles \citep{Donnan23}.
		 }
 \label{fig:first_sfr}
\end{figure}

\begin{figure}
\centering		\includegraphics[width=8cm]{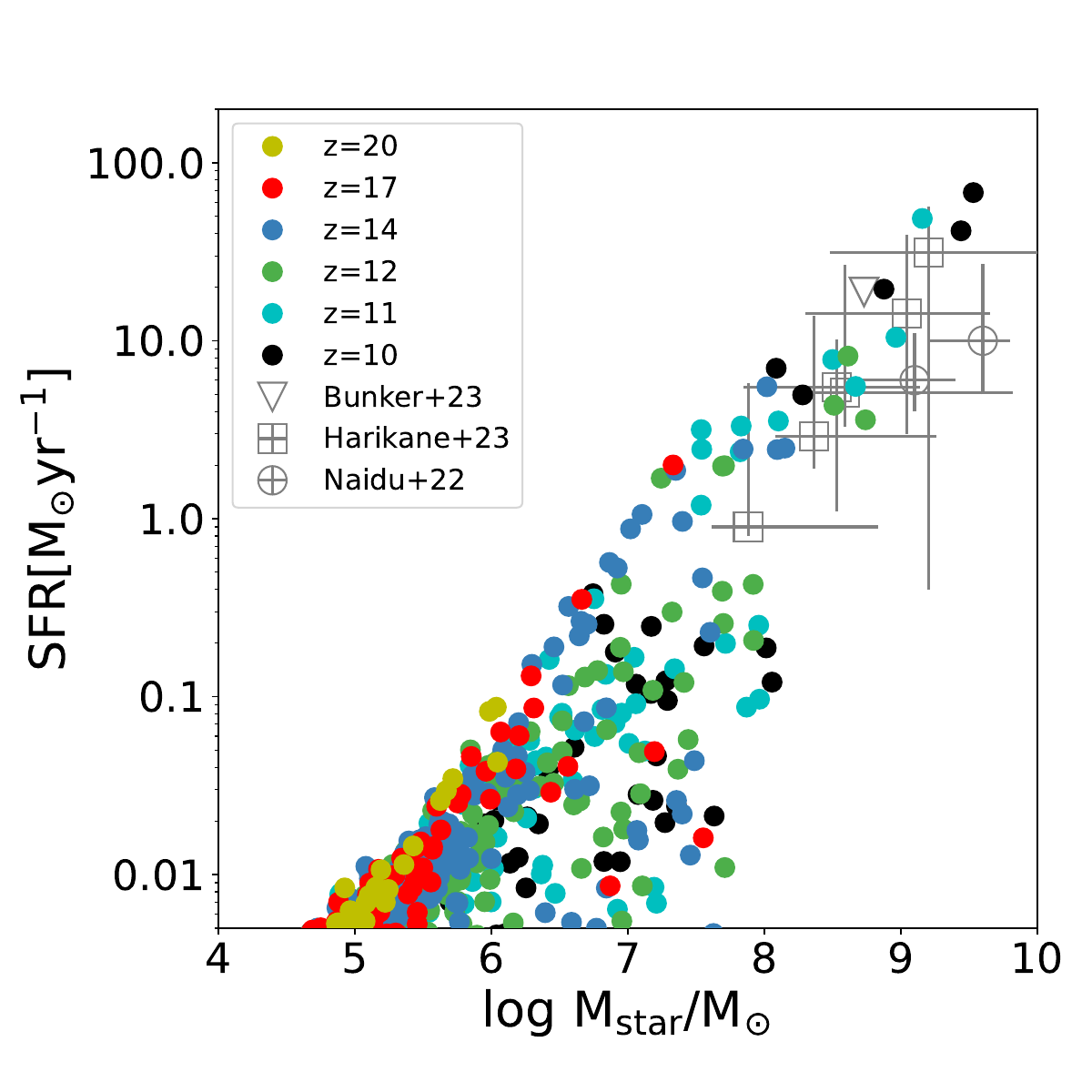}
	\caption{	
	Star formation rates as a function of stellar mass.
 Filled circles are our simulation results. Different colors  represent the different redshifts. Open symbols show the observational data: squares \citep{Harikane23}, circles \citep{Naidu22} and pentagons \citep{Bunker23}.
		 }
	\label{fig:sfr_mstar}
\end{figure}

Figure \ref{fig:2dmap} shows the column density maps of gas and stars. Stellar distributions are smoothed with a point spread function of JWST. The gas widely distributes within virial radii, while stellar distributions are compact and concentrated at the galactic centres. The gas accretes onto galaxies along the filamentary structures and the stellar feedback disturbs the gas structure. 
Stellar distributions and sizes change with time. At $z=12$, stellar clumps distribute at $0.5~R_{\rm vir}$, which reflects the minor merger phase. As the galaxy grows via baryon accretion, the size of the stellar components increases, but becomes small rapidly when major mergers happen \citep[see also,][]{Ono22}. 
Note that the size shrinkage after the merger process sensitively depends on the gas fraction and the structure of progenitor galaxies \citep[e.g.,][]{Dekel06b}.

Figure \ref{fig:first_sfr} presents the redshift evolution of halo, stellar masses, and SFR. 
The halo masses of the main progenitors are $\sim 10^{9}~\rm \Msun$ at $z \sim 20$ and evolve to $\sim 10^{11.5} ~\rm \Msun$ at $z \sim 10$. The fluctuations are due to mergers and the ability of the FOF group finder to identify all member particles.
The rarity of the halo with $10^{11.5}~\Msun$ at $z=10$ is ${\rm dN/dlnM_{\rm h} \sim 2 \times 10^{-7}~\rm cMpc^{-3}}$.
 The cosmic volume to host such a massive halo in our simulations is $\sim 500 ~\rm cMpc^{3}$ that is similar to the volumes of photometric galaxy surveys with JWST \citep[e.g.,][]{Finkelstein23}.
Therefore, it can be reasonable to directly compare our simulations with JWST data.
Note that, the rarity changes with time even for the most massive progenitors in the same region because of the variety of the halo merger history. 

As the halo grows, the stellar mass increases. The stellar masses of the main progenitors exceed $10^{8}~\rm \Msun$ at $z \sim 13 (14)$ and finally reach $4.3 (6.8) \times 10^{9}~\Msun$ in First0 (First1) run. The stellar masses are similar to observed galaxies at $z \lesssim 12$. On the other hand, it is much lower than the observed ones at $z \sim 16$. Suppose the estimated stellar masses of observed candidates are accurate and the redshifts are actually $\sim 16$. In that case, most gas is very efficiently converted into stars even in the early Universe \citep{Harikane23}.
\citet{Inayoshi22} suggested that $0.1-0.3$ of the gas should be converted into stars by using the abundance matching technique with the observed UV luminosity functions. 
In our simulations, the SN feedback efficiently works in the suppression of star formation. As a result, the gas is gradually converted into stars, and the star formation efficiency $\rm (M_{\rm star}/M_{\rm gas})$ of main progenitors in First0 run is $1.1\times10^{-1}, 2.0 \times 10^{-2}$ and $2.5 \times 10^{-2}$ at $z=10, 14$ and $17$. 
Note that, the redshifts and the physical properties of the candidate galaxies at $z>14$ were estimated with the photometric data. It is difficult to evaluate the impacts of emission lines only from the photometric data \citep[e.g.,][]{Schaerer09}. Therefore, their properties can be changed with follow-up spectroscopy \citep[e.g.,][]{ArrabalHaro23}.

The SFR also increases with the growth of  halo mass. Because of the cycle of suppression of star formation and the short recovery time scale of gas, the SFR fluctuates significantly with time \citep{Yajima17c}. Main progenitor galaxies can have $\gtrsim 10~\Msunyr$ at $z \lesssim 14$ and show starbursts with $\rm SFR = 50.4~ (133.5)~\Msunyr$ at $z=9.5$ in First0 (First1) run. If we consider the star formation rate as ${\rm SFR}=\epsilon \frac{M_{\rm gas}}{\tau_{\rm dyn}}$ where $\epsilon$ is an efficiency parameter and $\tau_{\rm dyn}$ is the dynamical time, the starbursts  at $z \sim 9.5$ correspond to $\epsilon \sim 0.08-0.25$. 
This value is much larger than typical star-forming galaxies in the local Universe. 
In the last period of $0.1~\rm Gyr$ ($z=9.5-11.5$), the halo mass of the main progenitors in First0 run increases from $1.2\times10^{11}~\Msun$ to $4.8 \times 10^{11}~\Msun$. The rapid mass growth with major mergers can induce the starburst. 
The SFRs at $z \lesssim 14$ nicely match with the observed ones by JWST \citep{Donnan23, Naidu22, Harikane23} and GN-z11 at $z=10.957$ \citep{Jiang21}.
Also, the modeled galaxies reproduce the observed $M_{\rm UV}$ at those redshifts.
As suggested in \citet{Yajima17c}, once the halo mass exceeds $10^{11}~\Msun$, most gas can be trapped in the deep gravitational potential against SN feedback. This can induce the starburst with $\rm SFR \gtrsim 10~\Msunyr$ and make galaxies observable.
In the redshift range, black holes are still in the state of being initial seeds with $\sim 10^{5}~\Msun$ and the accretion rates are mostly much lower than the Eddington limit. Therefore,  AGN feedback is negligible.  

Figure~\ref{fig:sfr_mstar} presents SFRs as a function of stellar mass. The SFR increases with the stellar mass. At $\Mstar \lesssim 10^{8}~\Msun$, the suppression of SFR due to the feedback makes the large dispersion. On the other hand, massive galaxies keep the star formation continuously.
We confirm that our results  match the observed galaxies. 
 The stellar mass monotonically increases with the halo mass although there is a large dispersion. The ratios of stellar to halo mass are $\sim 10^{-3}$ and $\sim 10^{-2}$ for $\Mh = 10^{10}$ and $10^{11}~\Msun$ at $z=10$. These values are similar to the empirical models in {\sc universemachine} project \citet{Behroozi20}, while it is somewhat higher than {\sc MilleniumTNG} \citep{Kannan22b}.
The conversion efficiency sensitively depends on the resolution, the star formation, and the feedback models. In particular, our simulations can resolve mini-haloes and dwarf galaxies, and their star formation. At high redshifts, stars formed in dwarf galaxies can contribute to the stellar mass in more massive galaxies via frequent merger processes. 

\begin{figure}
\centering		\includegraphics[width=8cm]{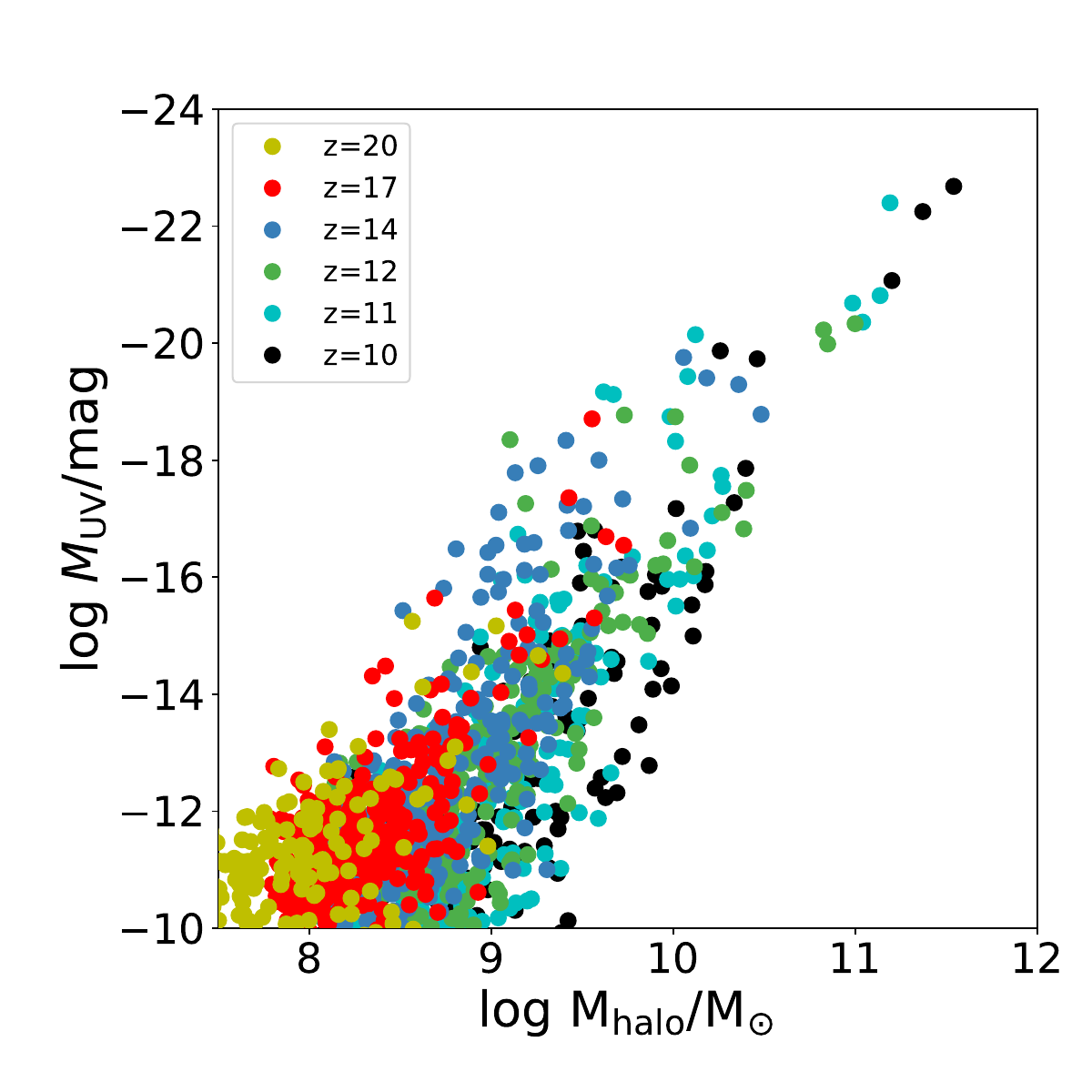}
\centering        \includegraphics[width=8cm]{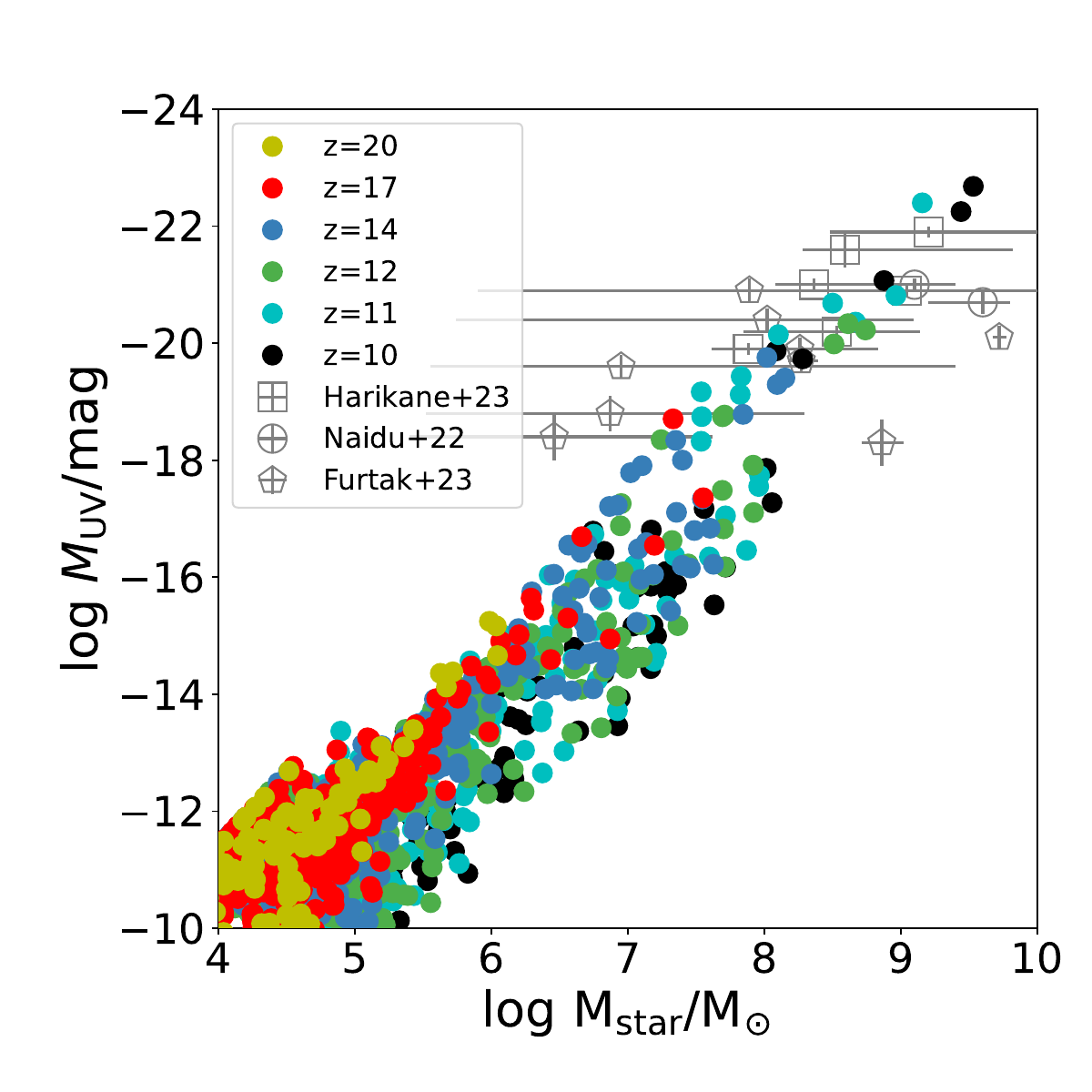}
	\caption{		
	UV fluxes as a function of halo and stellar masses.
 Filled circles are our simulation results. Different colors  represent the different redshifts. Open symbols show the observational data: squares \citep{Harikane23}, circles \citep{Naidu22} and pentagons \citep{Furtak23}.  
		 }
	\label{fig:Muv}
\end{figure}

\begin{figure}
\centering		\includegraphics[width=8cm]{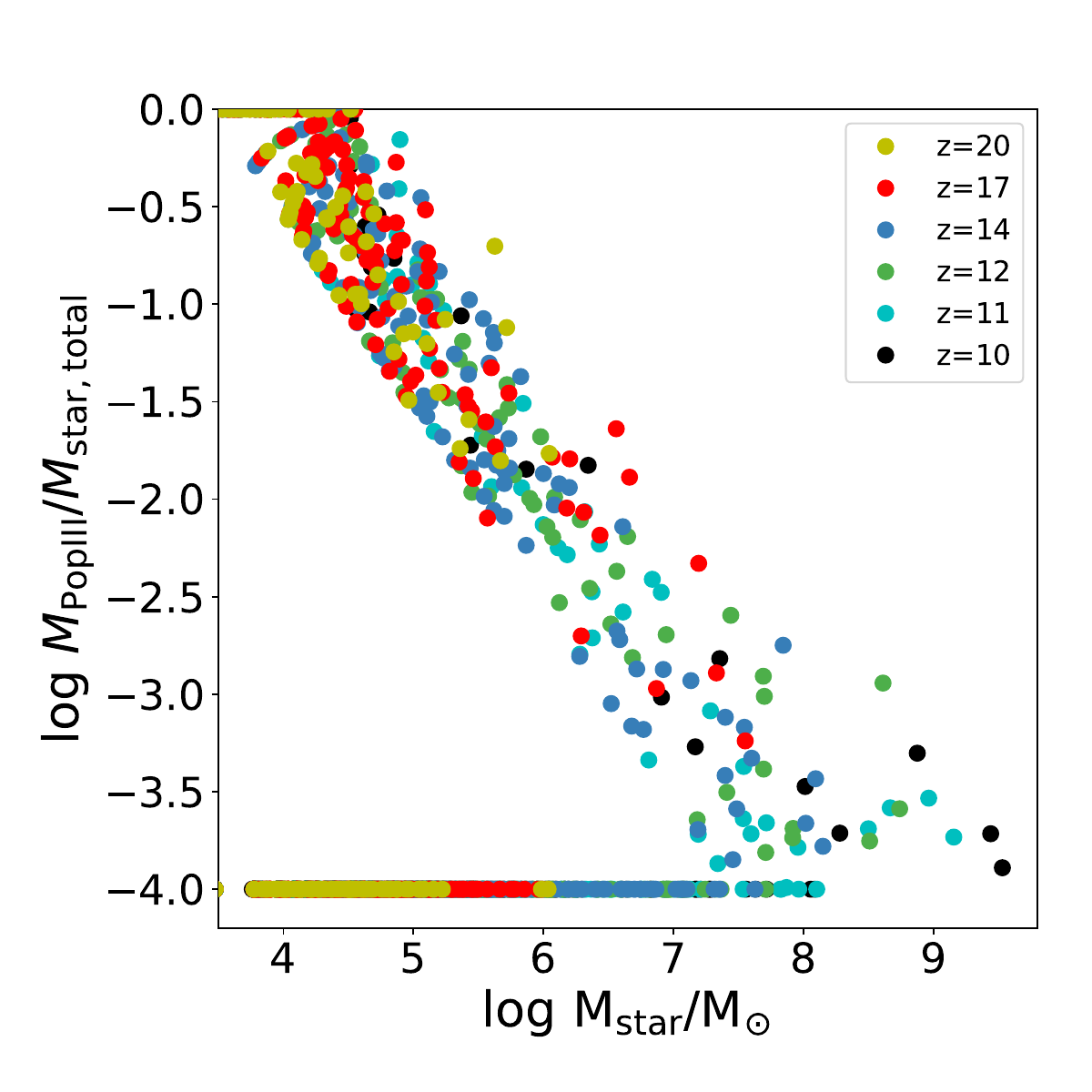}
\centering        \includegraphics[width=8cm]{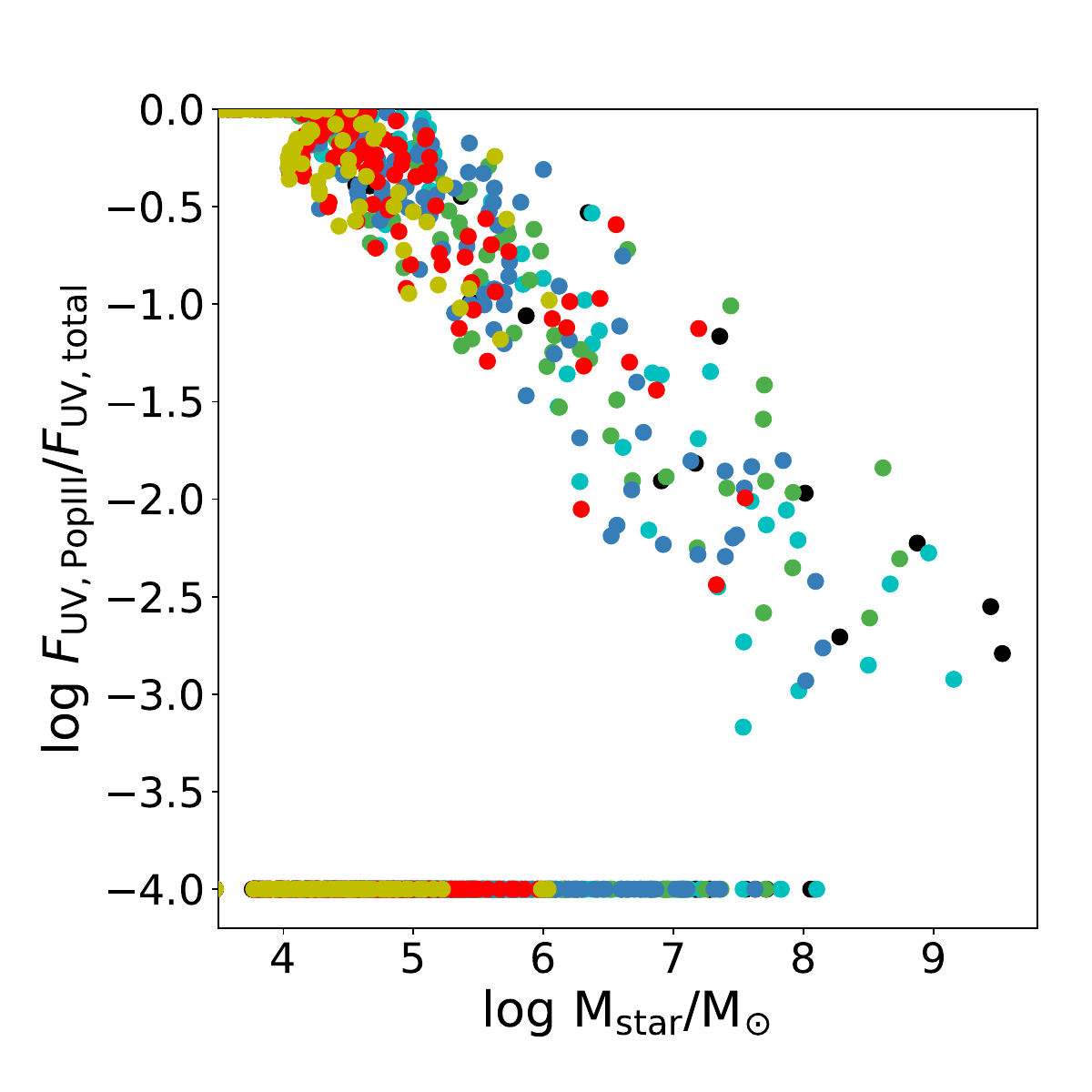}
	\caption{		
	Mass fraction of young population III stars to the total stellar mass (upper panel) and their contribution fraction to UV flux at $\lambda=1500~\rm \AA$ in the rest frame (lower panel).
 Different colors represent the different redshifts.
    The galaxies with mass fractions lower than $10^{-4}$ are plotted at $-4.0$.
		 }
	\label{fig:fpopiii}
\end{figure}

\begin{figure}
 \centering \includegraphics[width=9cm]{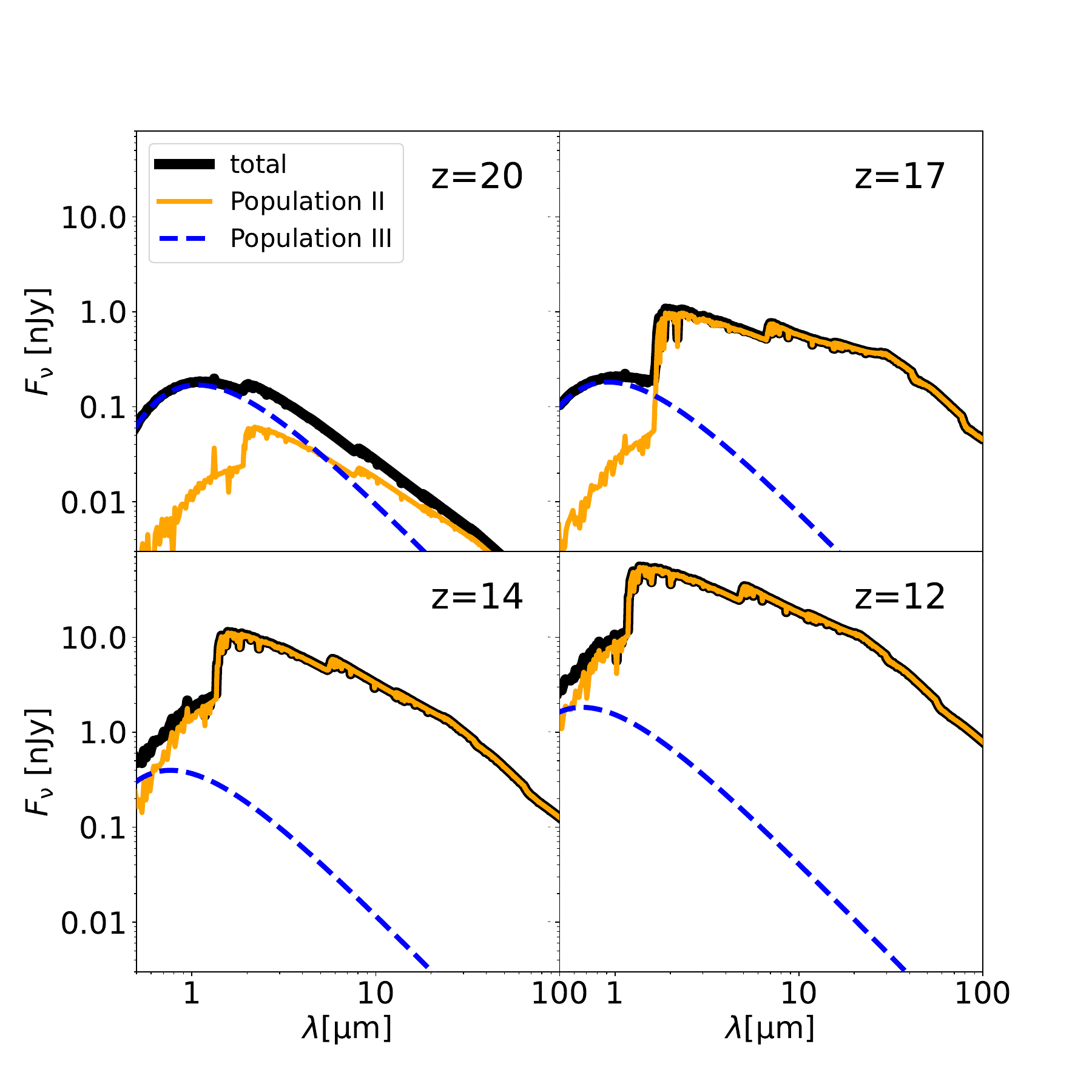}
	\caption{		
	SEDs of the most massive progenitors at $z=12, 14, 17$, and $20$. Blue dashed and orange solid lines show the radiation only from population III and II stars, respectively. Black solid lines are SEDs considering both stellar populations.
		 }
	\label{fig:first_sed}
\end{figure}

We present the relationships between the UV flux $\Muv$ and the halo and the stellar mass in Figure~\ref{fig:Muv}.
We estimate $\Muv$ by measuring the mean UV flux densities at $\lambda = 1500 - 2000~\rm \AA$ in modelled SEDs which will be shown in Figure~\ref{fig:first_sed}.
$\Muv$ is tightly related to the SFR although it somewhat changes depending on the star formation history. 
As the halo mass increases, the SFR becomes higher, resulting in the formation of bright galaxies.  
We find that the brightness can exceed the observable level $\Muv \lesssim -18~\rm mag$ if the halo mass is larger than $\sim 10^{10}~\Msun$. 
In low-mass haloes, galaxies at higher redshifts form stars more efficiently because they are compact and have higher gas density typically. Also, there is a large dispersion. This can be due to the SN feedback that induces the intermittent star formation history via the cycle of gas inflow and outflow \citep{Yajima17c}. On the other hand, the UV brightness is more tightly correlated with the stellar mass. 
Our simulations reproduce the observed UV brightnesses nicely.
Galaxies with $\Mstar \gtrsim 10^{8}~\Msun$ are likely to have  observable UV brightness $\Muv \lesssim -18~\rm mag$. 
As shown in Figure~\ref{fig:first_sfr}, the SFR rapidly increases as the halo mass increases. Therefore, the stellar masses in the massive haloes can be contributed mainly by the current starburst episode, resulting in the tight relation in the massive systems. 
 Note that some observed galaxies with $\Mstar \sim 10^{6-7}~\Msun$ are brighter than our modelled galaxies although they are within the error bars. As one possibility, hidden faint AGNs might contribute to observed UV fluxes \citep[e.g.,][]{Bunker23}. Future deep spectroscopic studies will allow us to investigate AGN activities.

An upper panel of figure~\ref{fig:fpopiii} shows the mass fraction of young population III stars to the total stellar mass. 
As the star formation proceeds, the interstellar gas is metal-enriched via type-II supernovae. 
Therefore, the fraction steeply decreases as the stellar mass increases. 
Also, some fractions of galaxies have no young Pop III stars. This indicates Pop III stars form only when primordial gas clouds accrete on a galaxy. 
Once the stellar mass exceeds $\sim 10^{7}~\Msun$, the fraction becomes $\lesssim 0.01$. 
Considering the sensitivities of current telescopes, only massive galaxies with $\Mstar \gtrsim 10^{7}~\Msun$ have been observed. Therefore, population II stars mainly form in the observed candidate galaxies at $z \gtrsim 10$.    
We find that the mass fraction is insensitive to the redshift in the range of $z=10-20$. Given that the metal production source is only type-II supernovae, the total metal mass released is simply proportional to the stellar mass. Thus, the insensitive redshift dependence indicates similar metal mixing with the interstellar gas in the redshift range. 
Our simulations suggest that low-mass galaxies with $\Mstar \lesssim 10^{5}~\Msun$ host population III stars with non-negligible fraction $\gtrsim 0.1$. 
 Recently, \citet{Riaz22} showed the mass fraction of Pop III stars to the total stellar masses based on their semi-numerical models \citep{Hartwig22}. The mass fraction of low-mass galaxies with $\Mstar \sim 10^{4-5}~\Msun$ is similar to their results. On the other hand, our results for massive galaxies are much higher. 
Our cosmological simulations indicate that the gas in mini-haloes can survive as the primordial state and contribute to the Pop III star formation in massive galaxies.

A lower panel of figure~\ref{fig:fpopiii} represents the contribution of Pop III stars to the UV flux at $\lambda=1500~\rm \AA$ in a rest frame. The contribution fraction also decreases with the mass of Pop III stars. However, since the mass-to-light ratio of Pop III stars is large and their effective temperature is high $T \sim 10^{5}~\rm K$ \citep{Schaerer02}, the contribution is moderately large even if the mass fraction of Pop III stars is low. In the cases of $\Mstar \sim 10^{7}~\Msun$, it can be a few percent. Therefore, a part of the observed fluxes by JWST could be contributed by PopIII stars. The UV flux can be dominated by Pop III stars if the stellar mass is lower than $\sim 10^{5}~\Msun$. However,
the low-mass systems are likely to be too faint for the sensitivities of current telescopes. Therefore, next-generation telescopes or  gravitationally lensed galaxies by foreground sources might be required for direct observations of population III star clusters.
Very recently, \citet{Vanzella23} indicated a candidate of a Pop III star cluster with the mass of $\lesssim 10^{4}~\Msun$ with the gravitational lens effect.
The total metallicity even for low-mass galaxies with $\Mstar \lesssim 10^{5}~\Msun$ exceed $\sim 10^{-3}~Z/Z_{\rm \odot}$. Therefore, the formation of population III stars indicates inhomogeneous metal enrichment in a galaxy. 
The formation sites of population III stars are somewhat far from the high-density regions of population II stars where the metal enrichment proceeds earlier. 
Also, we find that the number fraction of galaxies hosting young Pop III stars increases from $\sim 0.3$ for $\Mstar=10^{5}~\Msun$ to $\sim 0.7$ for $\Mstar=10^{8}~\Msun$. 
The low-mass galaxies without young Pop III stars consist of two states, star-forming only with Pop II stars or quenching of star formation due to the SN feedback.
Massive haloes are likely to distribute near the centre of overdense regions and primordial gas hosted by mini-haloes can accrete them frequently. Therefore, the massive haloes can host young Pop III stars although the mass fraction is low. 

Note that, even for the population III stars, we model the star formation by replacing a gas particle with a stellar particle with the uniform mass $\sim 8 \times 10^{3}~\Msun$, which models a star cluster. Therefore, the stellar particles release the same SN energy and metal mass. However, if the total stellar mass is smaller than $\sim 10^{3}~\Msun$ in low-mass haloes, the IMF may not be universal, resulting in unequal SN feedback and metal amount \citep{Abe21}. This can enhance spatial fluctuation of metal distribution in the large cosmic volume. Therefore, the metal distributions in low-mass galaxies are likely to change with the resolution and the model of population III stars. We will investigate these impacts on the fraction of population III stars in future work. 

Here, we derive intrinsic SEDs of galaxies by using a stellar synthesis code {\sc starburst99} \citep{Leitherer99}. 
Figure~\ref{fig:first_sed} shows the SEDs of the most massive progenitors at $z=12, 14, 17$, and $20$. 
In this work, we estimate SEDs with the optically-thin approximation, i.e., no dust attenuation, which can be reasonable for low-mass and low-metallicity galaxies \citep{Yajima12b, Yajima14c, Cullen17}.  The observed blue UV-slope ($\beta \lesssim -2$) supports the assumption \citep{Furtak23, Naidu22}.
The contribution from population III stars is estimated with the assumption of the brightness temperature of $10^{5}~\rm K$ and the mass-to-luminosity ratio for $120~\Msun$ derived in \citet{Schaerer02}. 
The mass fractions of Pop III stars of the galaxies are $1.1 \times 10^{-3}, 1.8 \times 10^{-3}, 4.7 \times 10^{-3}, $ and $0.2$ at $z=12, 14, 17,$ and $20$, respectively.
Their contributions to UV luminosity densities at $\lambda=1500~\rm \AA$ are $1.4 \times 10^{-2}, 1.6 \times 10^{-2}, 7.5 \times 10^{-2}$ 
and $0.57$. 
We find that the contribution fraction can be fit by ${\rm log_{10}}f_{\rm UV, PopIII} = 0.66 \; {\rm log_{10}} \left( \frac{M_{\rm star, PopIII}}{\Mstar}\right) + 0.20$. 
At $z=20$, the light from Pop III stars dominates at UV wavelengths. At the lower redshifts, the UV continuum fluxes are dominated by Pop II stars. 
We measure the UV slopes of the SEDs by using the flux densities at $\lambda=1300, 2000$, and $3000~\rm \AA$. It shows $\beta=-2.68, -2.76, -2.62$ and $-3.33$ at $z=12,14,17$ and $20$, respectively. These naturally reproduce the very blue slopes of the observed galaxies \citep{Atek22, Furtak23}.
Note that, we do not consider the nebular emission in the above SEDs. If the nebular continuum at the UV wavelengths is added into the SEDs, the slopes are changed to $\beta=-2.55, -2.60, -2.57$ and $-2.98$ at $z=12,14,17$ and $20$.
Furthermore, dust extinction can make the SEDs redder at  lower redshifts. 

The galaxies at $z\le 14$ have UV flux densities of $\gtrsim 10~\rm nJy$ that are observable by JWST with a reasonable integration time. We suggest that a part of the observed fluxes of candidate galaxies at $z \gtrsim 10$ could be contributed by Pop III stars. 
If the sensitivity of future observations at $3~\rm \micron$ will reach $\sim 0.1~\rm nJy$, the UV light dominated by Pop III stars can be observed directly.
Also, note that Lyman continuum fluxes can be dominated by Pop III stars significantly even if the mass fraction of Pop III stars is low $\lesssim 0.1$. For example, the contribution of Pop III stars to the LyC flux at $900~\rm \AA$ for the galaxy at $z=17$ is $0.7$.
Thus, these galaxies consisting of both Pop II and III stars can be strong ionizing sources. In addition, they may have unique properties in SEDs with high equivalent widths of doubly ionized oxygen, carbon, and helium \citep[see also,][]{Nakajima22}.
In this work, we do not take into account the radiative transfer in the galaxies. However, since the galaxies are low-mass and low-metallicity systems, non-ionizing UV continuum photons are expected to escape efficiently. Therefore, the estimated UV-slopes are unlikely to change significantly. On the other hand, the escape fraction of LyC photons can change with time depending on the inhomogeneous gas structure due to the SN feedback \citep{Yajima14c, Paardekooper15, Kimm15, Trebitsch17}.  In the case of a high escape fraction, He{\sc ii} and other metal lines can be faint.  As shown in \citet{Ono22}, the sizes of our modelled galaxies change with time significantly. In phases when dusty gas compactly distributes star-forming regions at the galactic center, UV photons can be attenuated even at such high redshifts. 
In practice, the gas at the galactic center reaches a metallicity with $Z \gtrsim 0.1~\Zsun$ at $z \sim 10$ (see also, Isobe et al. 2023b).
The escape fraction of photons sensitively depends on the covering fraction of dusty gas clouds from young stars. We will perform radiative transfer simulations in future work.

%
%



%
%

\section{Discussion \& Summary}
We have investigated the star formation and physical properties in the first galaxies formed in overdense regions modelled by the {\sc forever22} simulation project. 
Our simulations followed the evolution from mini-haloes hosting Pop III stars to massive galaxies with $\Mh > 10^{11}~\Msun$. Our findings are summarized as follows. 

\begin{itemize}
    \item SFR increases with the halo mass and changes in the short-time period due to the supernova feedback. Once the halo mass exceeds $\sim 10^{11}~\Msun$, galaxies continuously form stars with $\rm SFR \gtrsim 10~\Msunyr$ and induce starbursts with $\sim 100~\Msunyr$. Even massive galaxies in overdense regions cannot reproduce the observed stellar masses and SFRs of candidate galaxies at $z \sim 16$ suggested by \citet{Donnan23}, \citet{Harikane23} and \citet{Naidu22}.
    \item Our simulations reproduce the relation between $\Muv$ and $\Mstar$ of the observed galaxies at $z \gtrsim 10$ nicely. The galaxies with $\Mstar \gtrsim 10^{8}~\Msun$ show UV brightness of $\lesssim - 18~\rm mag$ which is observable by JWST. 
    \item Even when the galaxy is metal enriched and forms Pop II stars, Pop III stars can form in zero-metallicity spots. The mass fraction of Pop III stars decreases as the stellar mass increases, and it is $< 0.01$ for galaxies with $\Mstar \sim 10^{7}~\Msun$. Therefore, candidate galaxies at $z \gtrsim 10$ by JWST can be dominated by Pop II stars. We suggest that a part of galaxies with $\Mstar \lesssim 10^{5}~\Msun$ can host Pop III stars with a non-negligible fraction $\gtrsim 0.1$. 
    \item We model SEDs of galaxies with Pop II and Pop III stars. The UV continuum fluxes of massive galaxies at $z \le 17$ are dominated by Pop II stars.
    However, a few percent of UV fluxes can be from Pop III stars because of their large mass-to-luminosity ratio.
    The galaxies at $z \le 14$ have the brightness of $\gtrsim 10~\rm nJy$ at $\lambda=2~\rm \micron$ which can be observable by JWST with a reasonable integration time. 
\end{itemize}

The estimated physical properties and redshifts of the observed galaxies at $z\gtrsim 10$ are not robust. Future spectroscopic observations would present more reliable data and constrain the physical properties of galaxies. On the other hand, the physical properties of first galaxies modelled by numerical simulations can depend on the resolution and the models of star formation and feedback \citep{Abe21}. Also, the seeding of the first massive black holes is still under debate \citep{Inayoshi20}, and it may change the physical properties of galaxies and SEDs. 
 The star formation efficiency of massive haloes with $\Mh \gtrsim 10^{10}~\Msun$ at $z \sim 10$ is similar to the results in \citet{Behroozi20} and somewhat higher than \citet{Kannan22b}. In this paper, we newly provide insights about the relationship between Pop III stars and massive galaxies at $z \gtrsim 10$.  The physical properties are likely to sensitively depend on the resolution, the star formation, and the feedback models.
We will investigate the model and resolution dependencies on the first galaxy formation in future work. 

%
%
\section*{Acknowledgments}

 The numerical simulations were performed on the computer cluster, XC50 in NAOJ, and Trinity at Center for Computational Sciences in University of Tsukuba. This work is supported in part by MEXT/JSPS KAKENHI Grant Number 17H04827, 20H04724, 21H04489 (HY), 20K22358, 22H01258 (TH), NAOJ ALMA Scientific Research Grant Numbers 2019-11A, JST FOREST Program, Grant Number JP-MJFR202Z, and Astro Biology Center Project research AB041008 (HY). 
For the purpose of open access, the author has applied a Creative Commons Attribution (CC BY) licence to any Author Accepted Manuscript version arising from this submission.

\section*{Data availability}

The data underlying this article will be shared on reasonable request to the corresponding author.

%
%

\bibliographystyle{mnras}
\bibliography{HY} 

\begin{thebibliography}{}
\makeatletter
\relax
\def\mn@urlcharsother{\let\do\@makeother \do\$\do\&\do\#\do\^\do\_\do\%\do\~}
\def\mn@doi{\begingroup\mn@urlcharsother \@ifnextchar [ {\mn@doi@}
  {\mn@doi@[]}}
\def\mn@doi@[#1]#2{\def\@tempa{#1}\ifx\@tempa\@empty \href
  {http://dx.doi.org/#2} {doi:#2}\else \href {http://dx.doi.org/#2} {#1}\fi
  \endgroup}
\def\mn@eprint#1#2{\mn@eprint@#1:#2::\@nil}
\def\mn@eprint@arXiv#1{\href {http://arxiv.org/abs/#1} {{\tt arXiv:#1}}}
\def\mn@eprint@dblp#1{\href {http://dblp.uni-trier.de/rec/bibtex/#1.xml}
  {dblp:#1}}
\def\mn@eprint@#1:#2:#3:#4\@nil{\def\@tempa {#1}\def\@tempb {#2}\def\@tempc
  {#3}\ifx \@tempc \@empty \let \@tempc \@tempb \let \@tempb \@tempa \fi \ifx
  \@tempb \@empty \def\@tempb {arXiv}\fi \@ifundefined
  {mn@eprint@\@tempb}{\@tempb:\@tempc}{\expandafter \expandafter \csname
  mn@eprint@\@tempb\endcsname \expandafter{\@tempc}}}

\bibitem[\protect\citeauthoryear{{Abe}, {Yajima}, {Khochfar}, {Dalla Vecchia}
  \& {Omukai}}{{Abe} et~al.}{2021}]{Abe21}
{Abe} M.,  {Yajima} H.,  {Khochfar} S.,  {Dalla Vecchia} C.,   {Omukai} K.,
  2021, \mn@doi [\mnras] {10.1093/mnras/stab2637}, \href
  {https://ui.adsabs.harvard.edu/abs/2021MNRAS.508.3226A} {508, 3226}

\bibitem[\protect\citeauthoryear{{Agarwal}, {Dalla Vecchia}, {Johnson},
  {Khochfar}  \& {Paardekooper}}{{Agarwal} et~al.}{2014}]{Agarwal14}
{Agarwal} B.,  {Dalla Vecchia} C.,  {Johnson} J.~L.,  {Khochfar} S.,
  {Paardekooper} J.-P.,  2014, \mn@doi [\mnras] {10.1093/mnras/stu1112}, \href
  {http://adsabs.harvard.edu/abs/2014MNRAS.443..648A} {443, 648}

\bibitem[\protect\citeauthoryear{{Arata}, {Yajima}, {Nagamine}, {Li}  \&
  {Khochfar}}{{Arata} et~al.}{2019}]{Arata19}
{Arata} S.,  {Yajima} H.,  {Nagamine} K.,  {Li} Y.,   {Khochfar} S.,  2019,
  \mn@doi [\mnras] {10.1093/mnras/stz1887}, \href
  {https://ui.adsabs.harvard.edu/abs/2019MNRAS.488.2629A} {488, 2629}

\bibitem[\protect\citeauthoryear{{Arrabal Haro} et~al.,}{{Arrabal Haro}
  et~al.}{2023}]{ArrabalHaro23}
{Arrabal Haro} P.,  et~al., 2023, \mn@doi [arXiv e-prints]
  {10.48550/arXiv.2303.15431}, \href
  {https://ui.adsabs.harvard.edu/abs/2023arXiv230315431A} {p. arXiv:2303.15431}

\bibitem[\protect\citeauthoryear{{Atek} et~al.,}{{Atek} et~al.}{2022}]{Atek22}
{Atek} H.,  et~al., 2022, arXiv e-prints, \href
  {https://ui.adsabs.harvard.edu/abs/2022arXiv220712338A} {p. arXiv:2207.12338}

\bibitem[\protect\citeauthoryear{{Bakx} et~al.,}{{Bakx} et~al.}{2022}]{Bakx22}
{Bakx} T. J.~L.~C.,  et~al., 2022, arXiv e-prints, \href
  {https://ui.adsabs.harvard.edu/abs/2022arXiv220813642B} {p. arXiv:2208.13642}

\bibitem[\protect\citeauthoryear{{Behroozi}, {Wechsler}, {Hearin}  \&
  {Conroy}}{{Behroozi} et~al.}{2019}]{Behroozi19}
{Behroozi} P.,  {Wechsler} R.~H.,  {Hearin} A.~P.,   {Conroy} C.,  2019,
  \mn@doi [\mnras] {10.1093/mnras/stz1182}, \href
  {https://ui.adsabs.harvard.edu/abs/2019MNRAS.488.3143B} {488, 3143}

\bibitem[\protect\citeauthoryear{{Behroozi} et~al.,}{{Behroozi}
  et~al.}{2020}]{Behroozi20}
{Behroozi} P.,  et~al., 2020, \mn@doi [\mnras] {10.1093/mnras/staa3164}, \href
  {https://ui.adsabs.harvard.edu/abs/2020MNRAS.499.5702B} {499, 5702}

\bibitem[\protect\citeauthoryear{{Bouwens}, {Stefanon}, {Oesch}, {Illingworth},
  {Nanayakkara}, {Roberts-Borsani}, {Labb{\'e}}  \& {Smit}}{{Bouwens}
  et~al.}{2019}]{Bouwens19}
{Bouwens} R.~J.,  {Stefanon} M.,  {Oesch} P.~A.,  {Illingworth} G.~D.,
  {Nanayakkara} T.,  {Roberts-Borsani} G.,  {Labb{\'e}} I.,   {Smit} R.,  2019,
  \mn@doi [\apj] {10.3847/1538-4357/ab24c5}, \href
  {https://ui.adsabs.harvard.edu/abs/2019ApJ...880...25B} {880, 25}

\bibitem[\protect\citeauthoryear{{Bunker} et~al.,}{{Bunker}
  et~al.}{2023}]{Bunker23}
{Bunker} A.~J.,  et~al., 2023, \mn@doi [arXiv e-prints]
  {10.48550/arXiv.2302.07256}, \href
  {https://ui.adsabs.harvard.edu/abs/2023arXiv230207256B} {p. arXiv:2302.07256}

\bibitem[\protect\citeauthoryear{{Capak} et~al.,}{{Capak}
  et~al.}{2015}]{Capak15}
{Capak} P.~L.,  et~al., 2015, \mn@doi [\nat] {10.1038/nature14500}, \href
  {https://ui.adsabs.harvard.edu/abs/2015Natur.522..455C} {522, 455}

\bibitem[\protect\citeauthoryear{{Chiaki} \& {Wise}}{{Chiaki} \&
  {Wise}}{2019}]{Chiaki19}
{Chiaki} G.,  {Wise} J.~H.,  2019, \mn@doi [\mnras] {10.1093/mnras/sty2984},
  \href {https://ui.adsabs.harvard.edu/abs/2019MNRAS.482.3933C} {482, 3933}

\bibitem[\protect\citeauthoryear{{Chon}, {Omukai}  \& {Schneider}}{{Chon}
  et~al.}{2021}]{Chon21}
{Chon} S.,  {Omukai} K.,   {Schneider} R.,  2021, \mn@doi [\mnras]
  {10.1093/mnras/stab2497}, \href
  {https://ui.adsabs.harvard.edu/abs/2021MNRAS.508.4175C} {508, 4175}

\bibitem[\protect\citeauthoryear{{Cullen}, {McLure}, {Khochfar}, {Dunlop}  \&
  {Dalla Vecchia}}{{Cullen} et~al.}{2017}]{Cullen17}
{Cullen} F.,  {McLure} R.~J.,  {Khochfar} S.,  {Dunlop} J.~S.,   {Dalla
  Vecchia} C.,  2017, \mn@doi [\mnras] {10.1093/mnras/stx1451}, \href
  {https://ui.adsabs.harvard.edu/abs/2017MNRAS.470.3006C} {470, 3006}

\bibitem[\protect\citeauthoryear{{Cullen} et~al.,}{{Cullen}
  et~al.}{2022}]{Cullen22}
{Cullen} F.,  et~al., 2022, arXiv e-prints, \href
  {https://ui.adsabs.harvard.edu/abs/2022arXiv220804914C} {p. arXiv:2208.04914}

\bibitem[\protect\citeauthoryear{{Dalla Vecchia} \& {Schaye}}{{Dalla Vecchia}
  \& {Schaye}}{2012}]{DallaVecchia12}
{Dalla Vecchia} C.,  {Schaye} J.,  2012, \mn@doi [\mnras]
  {10.1111/j.1365-2966.2012.21704.x}, \href
  {http://adsabs.harvard.edu/abs/2012MNRAS.426..140D} {426, 140}

\bibitem[\protect\citeauthoryear{{Dekel} \& {Cox}}{{Dekel} \&
  {Cox}}{2006}]{Dekel06b}
{Dekel} A.,  {Cox} T.~J.,  2006, \mn@doi [\mnras]
  {10.1111/j.1365-2966.2006.10566.x}, \href
  {https://ui.adsabs.harvard.edu/abs/2006MNRAS.370.1445D} {370, 1445}

\bibitem[\protect\citeauthoryear{{Donnan} et~al.,}{{Donnan}
  et~al.}{2023}]{Donnan23}
{Donnan} C.~T.,  et~al., 2023, \mn@doi [\mnras] {10.1093/mnras/stac3472}, \href
  {https://ui.adsabs.harvard.edu/abs/2023MNRAS.518.6011D} {518, 6011}

\bibitem[\protect\citeauthoryear{{Ferland}}{{Ferland}}{2000}]{Ferland00}
{Ferland} G.~J.,  2000, in {Arthur} S.~J.,  {Brickhouse} N.~S.,   {Franco} J.,
  eds,  Revista Mexicana de Astronomia y Astrofisica Conference Series Vol. 9,
  Revista Mexicana de Astronomia y Astrofisica Conference Series. pp 153--157

\bibitem[\protect\citeauthoryear{{Finkelstein} et~al.,}{{Finkelstein}
  et~al.}{2013}]{Finkelstein13}
{Finkelstein} S.~L.,  et~al., 2013, \mn@doi [\nat] {10.1038/nature12657}, \href
  {http://adsabs.harvard.edu/abs/2013Natur.502..524F} {502, 524}

\bibitem[\protect\citeauthoryear{{Finkelstein} et~al.,}{{Finkelstein}
  et~al.}{2023}]{Finkelstein23}
{Finkelstein} S.~L.,  et~al., 2023, \mn@doi [\apjl] {10.3847/2041-8213/acade4},
  \href {https://ui.adsabs.harvard.edu/abs/2023ApJ...946L..13F} {946, L13}

\bibitem[\protect\citeauthoryear{{Frebel}, {Johnson}  \& {Bromm}}{{Frebel}
  et~al.}{2007}]{Frebel07}
{Frebel} A.,  {Johnson} J.~L.,   {Bromm} V.,  2007, \mn@doi [\mnras]
  {10.1111/j.1745-3933.2007.00344.x}, \href
  {https://ui.adsabs.harvard.edu/abs/2007MNRAS.380L..40F} {380, L40}

\bibitem[\protect\citeauthoryear{{Freedman}}{{Freedman}}{2021}]{Freedman21}
{Freedman} W.~L.,  2021, \mn@doi [\apj] {10.3847/1538-4357/ac0e95}, \href
  {https://ui.adsabs.harvard.edu/abs/2021ApJ...919...16F} {919, 16}

\bibitem[\protect\citeauthoryear{{Furtak}, {Shuntov}, {Atek}, {Zitrin},
  {Richard}, {Lehnert}  \& {Chevallard}}{{Furtak} et~al.}{2023}]{Furtak23}
{Furtak} L.~J.,  {Shuntov} M.,  {Atek} H.,  {Zitrin} A.,  {Richard} J.,
  {Lehnert} M.~D.,   {Chevallard} J.,  2023, \mn@doi [\mnras]
  {10.1093/mnras/stac3717}, \href
  {https://ui.adsabs.harvard.edu/abs/2023MNRAS.519.3064F} {519, 3064}

\bibitem[\protect\citeauthoryear{{Gabrielpillai}, {Somerville}, {Genel},
  {Rodriguez-Gomez}, {Pandya}, {Yung}  \& {Hernquist}}{{Gabrielpillai}
  et~al.}{2022}]{Gabrielpillai22}
{Gabrielpillai} A.,  {Somerville} R.~S.,  {Genel} S.,  {Rodriguez-Gomez} V.,
  {Pandya} V.,  {Yung} L.~Y.~A.,   {Hernquist} L.,  2022, \mn@doi [\mnras]
  {10.1093/mnras/stac2297}, \href
  {https://ui.adsabs.harvard.edu/abs/2022MNRAS.517.6091G} {517, 6091}

\bibitem[\protect\citeauthoryear{{Harikane} et~al.,}{{Harikane}
  et~al.}{2023}]{Harikane23}
{Harikane} Y.,  et~al., 2023, \mn@doi [\apjs] {10.3847/1538-4365/acaaa9}, \href
  {https://ui.adsabs.harvard.edu/abs/2023ApJS..265....5H} {265, 5}

\bibitem[\protect\citeauthoryear{{Hartwig} et~al.,}{{Hartwig}
  et~al.}{2022}]{Hartwig22}
{Hartwig} T.,  et~al., 2022, \mn@doi [\apj] {10.3847/1538-4357/ac7150}, \href
  {https://ui.adsabs.harvard.edu/abs/2022ApJ...936...45H} {936, 45}

\bibitem[\protect\citeauthoryear{{Hashimoto} et~al.,}{{Hashimoto}
  et~al.}{2018}]{Hashimoto18}
{Hashimoto} T.,  et~al., 2018, \mn@doi [\nat] {10.1038/s41586-018-0117-z},
  \href {https://ui.adsabs.harvard.edu/abs/2018Natur.557..392H} {557, 392}

\bibitem[\protect\citeauthoryear{{Hirano}, {Hosokawa}, {Yoshida}, {Omukai}  \&
  {Yorke}}{{Hirano} et~al.}{2015}]{Hirano15}
{Hirano} S.,  {Hosokawa} T.,  {Yoshida} N.,  {Omukai} K.,   {Yorke} H.~W.,
  2015, \mn@doi [\mnras] {10.1093/mnras/stv044}, \href
  {https://ui.adsabs.harvard.edu/abs/2015MNRAS.448..568H} {448, 568}

\bibitem[\protect\citeauthoryear{{Inayoshi}, {Visbal}  \& {Haiman}}{{Inayoshi}
  et~al.}{2020}]{Inayoshi20}
{Inayoshi} K.,  {Visbal} E.,   {Haiman} Z.,  2020, \mn@doi [\araa]
  {10.1146/annurev-astro-120419-014455}, \href
  {https://ui.adsabs.harvard.edu/abs/2020ARA&A..58...27I} {58, 27}

\bibitem[\protect\citeauthoryear{{Inayoshi}, {Harikane}, {Inoue}, {Li}  \&
  {Ho}}{{Inayoshi} et~al.}{2022}]{Inayoshi22}
{Inayoshi} K.,  {Harikane} Y.,  {Inoue} A.~K.,  {Li} W.,   {Ho} L.~C.,  2022,
  arXiv e-prints, \href {https://ui.adsabs.harvard.edu/abs/2022arXiv220806872I}
  {p. arXiv:2208.06872}

\bibitem[\protect\citeauthoryear{{Inoue} et~al.,}{{Inoue}
  et~al.}{2016}]{Inoue16}
{Inoue} A.~K.,  et~al., 2016, \mn@doi [Science] {10.1126/science.aaf0714},
  \href {https://ui.adsabs.harvard.edu/abs/2016Sci...352.1559I} {352, 1559}

\bibitem[\protect\citeauthoryear{{Jeon} \& {Bromm}}{{Jeon} \&
  {Bromm}}{2019}]{Jeon19}
{Jeon} M.,  {Bromm} V.,  2019, \mn@doi [\mnras] {10.1093/mnras/stz863}, \href
  {https://ui.adsabs.harvard.edu/abs/2019MNRAS.485.5939J} {485, 5939}

\bibitem[\protect\citeauthoryear{{Jiang} et~al.,}{{Jiang}
  et~al.}{2021}]{Jiang21}
{Jiang} L.,  et~al., 2021, \mn@doi [Nature Astronomy]
  {10.1038/s41550-020-01275-y}, \href
  {https://ui.adsabs.harvard.edu/abs/2021NatAs...5..256J} {5, 256}

\bibitem[\protect\citeauthoryear{{Johnson}, {Dalla}  \& {Khochfar}}{{Johnson}
  et~al.}{2013}]{Johnson13}
{Johnson} J.~L.,  {Dalla} V.~C.,   {Khochfar} S.,  2013, \mn@doi [\mnras]
  {10.1093/mnras/sts011}, \href
  {http://adsabs.harvard.edu/abs/2013MNRAS.428.1857J} {428, 1857}

\bibitem[\protect\citeauthoryear{{Kannan} et~al.,}{{Kannan}
  et~al.}{2022a}]{Kannan22b}
{Kannan} R.,  et~al., 2022a, \mn@doi [arXiv e-prints]
  {10.48550/arXiv.2210.10066}, \href
  {https://ui.adsabs.harvard.edu/abs/2022arXiv221010066K} {p. arXiv:2210.10066}

\bibitem[\protect\citeauthoryear{{Kannan}, {Garaldi}, {Smith}, {Pakmor},
  {Springel}, {Vogelsberger}  \& {Hernquist}}{{Kannan}
  et~al.}{2022b}]{Kannan22a}
{Kannan} R.,  {Garaldi} E.,  {Smith} A.,  {Pakmor} R.,  {Springel} V.,
  {Vogelsberger} M.,   {Hernquist} L.,  2022b, \mn@doi [\mnras]
  {10.1093/mnras/stab3710}, \href
  {https://ui.adsabs.harvard.edu/abs/2022MNRAS.511.4005K} {511, 4005}

\bibitem[\protect\citeauthoryear{{Kimm}, {Cen}, {Rosdahl}  \& {Yi}}{{Kimm}
  et~al.}{2015}]{Kimm15}
{Kimm} T.,  {Cen} R.,  {Rosdahl} J.,   {Yi} S.,  2015, preprint, \href
  {http://adsabs.harvard.edu/abs/2015arXiv151005671K} {} (\mn@eprint {arXiv}
  {1510.05671})

\bibitem[\protect\citeauthoryear{{Komatsu} et~al.,}{{Komatsu}
  et~al.}{2011}]{Komatsu11}
{Komatsu} E.,  et~al., 2011, \mn@doi [\apjs] {10.1088/0067-0049/192/2/18},
  \href {http://ads.nao.ac.jp/abs/2011ApJS..192...18K} {192, 18}

\bibitem[\protect\citeauthoryear{{Latif}, {Whalen}, {Khochfar}, {Herrington}
  \& {Woods}}{{Latif} et~al.}{2022a}]{Latif22b}
{Latif} M.~A.,  {Whalen} D.~J.,  {Khochfar} S.,  {Herrington} N.~P.,   {Woods}
  T.~E.,  2022a, \mn@doi [\nat] {10.1038/s41586-022-04813-y}, \href
  {https://ui.adsabs.harvard.edu/abs/2022Natur.607...48L} {607, 48}

\bibitem[\protect\citeauthoryear{{Latif}, {Whalen}  \& {Khochfar}}{{Latif}
  et~al.}{2022b}]{Latif22}
{Latif} M.~A.,  {Whalen} D.,   {Khochfar} S.,  2022b, \mn@doi [\apj]
  {10.3847/1538-4357/ac3916}, \href
  {https://ui.adsabs.harvard.edu/abs/2022ApJ...925...28L} {925, 28}

\bibitem[\protect\citeauthoryear{{Leitherer} et~al.,}{{Leitherer}
  et~al.}{1999}]{Leitherer99}
{Leitherer} C.,  et~al., 1999, \mn@doi [\apjs] {10.1086/313233}, \href
  {http://adsabs.harvard.edu/abs/1999ApJS..123....3L} {123, 3}

\bibitem[\protect\citeauthoryear{{Liu} \& {Bromm}}{{Liu} \&
  {Bromm}}{2020}]{Liu20}
{Liu} B.,  {Bromm} V.,  2020, \mn@doi [\mnras] {10.1093/mnras/staa2143}, \href
  {https://ui.adsabs.harvard.edu/abs/2020MNRAS.497.2839L} {497, 2839}

\bibitem[\protect\citeauthoryear{{Lovell}, {Vijayan}, {Thomas}, {Wilkins},
  {Barnes}, {Irodotou}  \& {Roper}}{{Lovell} et~al.}{2021}]{Lovell21}
{Lovell} C.~C.,  {Vijayan} A.~P.,  {Thomas} P.~A.,  {Wilkins} S.~M.,  {Barnes}
  D.~J.,  {Irodotou} D.,   {Roper} W.,  2021, \mn@doi [\mnras]
  {10.1093/mnras/staa3360}, \href
  {https://ui.adsabs.harvard.edu/abs/2021MNRAS.500.2127L} {500, 2127}

\bibitem[\protect\citeauthoryear{{Ma}, {Quataert}, {Wetzel}, {Hopkins},
  {Faucher-Gigu{\`e}re}  \& {Kere{\v{s}}}}{{Ma} et~al.}{2020}]{Ma20}
{Ma} X.,  {Quataert} E.,  {Wetzel} A.,  {Hopkins} P.~F.,  {Faucher-Gigu{\`e}re}
  C.-A.,   {Kere{\v{s}}} D.,  2020, \mn@doi [\mnras] {10.1093/mnras/staa2404},
  \href {https://ui.adsabs.harvard.edu/abs/2020MNRAS.498.2001M} {498, 2001}

\bibitem[\protect\citeauthoryear{{Maio}, {Ciardi}, {Dolag}, {Tornatore}  \&
  {Khochfar}}{{Maio} et~al.}{2010}]{Maio10}
{Maio} U.,  {Ciardi} B.,  {Dolag} K.,  {Tornatore} L.,   {Khochfar} S.,  2010,
  \mn@doi [\mnras] {10.1111/j.1365-2966.2010.17003.x}, \href
  {https://ui.adsabs.harvard.edu/abs/2010MNRAS.407.1003M} {407, 1003}

\bibitem[\protect\citeauthoryear{{Maio}, {Khochfar}, {Johnson}  \&
  {Ciardi}}{{Maio} et~al.}{2011}]{Maio11}
{Maio} U.,  {Khochfar} S.,  {Johnson} J.~L.,   {Ciardi} B.,  2011, \mn@doi
  [\mnras] {10.1111/j.1365-2966.2011.18455.x}, \href
  {http://adsabs.harvard.edu/abs/2011MNRAS.414.1145M} {414, 1145}

\bibitem[\protect\citeauthoryear{{Naidu} et~al.,}{{Naidu}
  et~al.}{2022}]{Naidu22}
{Naidu} R.~P.,  et~al., 2022, \mn@doi [\apjl] {10.3847/2041-8213/ac9b22}, \href
  {https://ui.adsabs.harvard.edu/abs/2022ApJ...940L..14N} {940, L14}

\bibitem[\protect\citeauthoryear{{Nakajima} \& {Maiolino}}{{Nakajima} \&
  {Maiolino}}{2022}]{Nakajima22}
{Nakajima} K.,  {Maiolino} R.,  2022, \mn@doi [\mnras]
  {10.1093/mnras/stac1242}, \href
  {https://ui.adsabs.harvard.edu/abs/2022MNRAS.513.5134N} {513, 5134}

\bibitem[\protect\citeauthoryear{{Ocvirk} et~al.,}{{Ocvirk}
  et~al.}{2016}]{Ocvirk16}
{Ocvirk} P.,  et~al., 2016, \mn@doi [\mnras] {10.1093/mnras/stw2036}, \href
  {http://adsabs.harvard.edu/abs/2016MNRAS.463.1462O} {463, 1462}

\bibitem[\protect\citeauthoryear{{Oesch} et~al.,}{{Oesch}
  et~al.}{2013}]{Oesch13}
{Oesch} P.~A.,  et~al., 2013, \mn@doi [\apj] {10.1088/0004-637X/773/1/75},
  \href {http://adsabs.harvard.edu/abs/2013ApJ...773...75O} {773, 75}

\bibitem[\protect\citeauthoryear{{Oesch} et~al.,}{{Oesch}
  et~al.}{2016}]{Oesch16}
{Oesch} P.~A.,  et~al., 2016, \mn@doi [\apj] {10.3847/0004-637X/819/2/129},
  \href {http://adsabs.harvard.edu/abs/2016ApJ...819..129O} {819, 129}

\bibitem[\protect\citeauthoryear{{Omukai}, {Tsuribe}, {Schneider}  \&
  {Ferrara}}{{Omukai} et~al.}{2005}]{Omukai05}
{Omukai} K.,  {Tsuribe} T.,  {Schneider} R.,   {Ferrara} A.,  2005, \mn@doi
  [\apj] {10.1086/429955}, \href
  {http://adsabs.harvard.edu/abs/2005ApJ...626..627O} {626, 627}

\bibitem[\protect\citeauthoryear{{Ono} et~al.,}{{Ono} et~al.}{2012}]{Ono12}
{Ono} Y.,  et~al., 2012, \mn@doi [\apj] {10.1088/0004-637X/744/2/83}, \href
  {http://ads.nao.ac.jp/abs/2012ApJ...744...83O} {744, 83}

\bibitem[\protect\citeauthoryear{{Ono} et~al.,}{{Ono} et~al.}{2022}]{Ono22}
{Ono} Y.,  et~al., 2022, arXiv e-prints, \href
  {https://ui.adsabs.harvard.edu/abs/2022arXiv220813582O} {p. arXiv:2208.13582}

\bibitem[\protect\citeauthoryear{{Ouchi} et~al.,}{{Ouchi}
  et~al.}{2018}]{Ouchi18}
{Ouchi} M.,  et~al., 2018, \mn@doi [\pasj] {10.1093/pasj/psx074}, \href
  {https://ui.adsabs.harvard.edu/abs/2018PASJ...70S..13O} {70, S13}

\bibitem[\protect\citeauthoryear{{Paardekooper}, {Khochfar}  \& {Dalla
  Vecchia}}{{Paardekooper} et~al.}{2013}]{Paardekooper13}
{Paardekooper} J.-P.,  {Khochfar} S.,   {Dalla Vecchia} C.,  2013, \mn@doi
  [\mnras] {10.1093/mnrasl/sls032}, \href
  {http://adsabs.harvard.edu/abs/2013MNRAS.429L..94P} {429, L94}

\bibitem[\protect\citeauthoryear{{Paardekooper}, {Khochfar}  \& {Dalla
  Vecchia}}{{Paardekooper} et~al.}{2015}]{Paardekooper15}
{Paardekooper} J.-P.,  {Khochfar} S.,   {Dalla Vecchia} C.,  2015, \mn@doi
  [\mnras] {10.1093/mnras/stv1114}, \href
  {http://adsabs.harvard.edu/abs/2015MNRAS.451.2544P} {451, 2544}

\bibitem[\protect\citeauthoryear{{Pakmor} et~al.,}{{Pakmor}
  et~al.}{2022}]{Pakmor22}
{Pakmor} R.,  et~al., 2022, \mn@doi [arXiv e-prints]
  {10.48550/arXiv.2210.10060}, \href
  {https://ui.adsabs.harvard.edu/abs/2022arXiv221010060P} {p. arXiv:2210.10060}

\bibitem[\protect\citeauthoryear{{Pallottini}, {Ferrara}, {Gallerani},
  {Salvadori}  \& {D'Odorico}}{{Pallottini} et~al.}{2014}]{Pallottini14}
{Pallottini} A.,  {Ferrara} A.,  {Gallerani} S.,  {Salvadori} S.,   {D'Odorico}
  V.,  2014, \mn@doi [\mnras] {10.1093/mnras/stu451}, \href
  {https://ui.adsabs.harvard.edu/abs/2014MNRAS.440.2498P} {440, 2498}

\bibitem[\protect\citeauthoryear{{Planck Collaboration} et~al.,}{{Planck
  Collaboration} et~al.}{2020}]{Planck20}
{Planck Collaboration} et~al., 2020, \mn@doi [\aap]
  {10.1051/0004-6361/201833910}, \href
  {https://ui.adsabs.harvard.edu/abs/2020A&A...641A...6P} {641, A6}

\bibitem[\protect\citeauthoryear{{Popping}}{{Popping}}{2022}]{Popping22}
{Popping} G.,  2022, arXiv e-prints, \href
  {https://ui.adsabs.harvard.edu/abs/2022arXiv220813072P} {p. arXiv:2208.13072}

\bibitem[\protect\citeauthoryear{{Regan} \& {Haehnelt}}{{Regan} \&
  {Haehnelt}}{2009}]{Regan09}
{Regan} J.~A.,  {Haehnelt} M.~G.,  2009, \mn@doi [\mnras]
  {10.1111/j.1365-2966.2009.14579.x}, \href
  {http://adsabs.harvard.edu/abs/2009MNRAS.396..343R} {396, 343}

\bibitem[\protect\citeauthoryear{{Riaz}, {Hartwig}  \& {Latif}}{{Riaz}
  et~al.}{2022}]{Riaz22}
{Riaz} S.,  {Hartwig} T.,   {Latif} M.~A.,  2022, \mn@doi [\apjl]
  {10.3847/2041-8213/ac8ea6}, \href
  {https://ui.adsabs.harvard.edu/abs/2022ApJ...937L...6R} {937, L6}

\bibitem[\protect\citeauthoryear{{Rosdahl} et~al.,}{{Rosdahl}
  et~al.}{2022}]{Rosdahl22}
{Rosdahl} J.,  et~al., 2022, \mn@doi [\mnras] {10.1093/mnras/stac1942}, \href
  {https://ui.adsabs.harvard.edu/abs/2022MNRAS.515.2386R} {515, 2386}

\bibitem[\protect\citeauthoryear{{Schaerer}}{{Schaerer}}{2002}]{Schaerer02}
{Schaerer} D.,  2002, \mn@doi [\aap] {10.1051/0004-6361:20011619}, \href
  {http://adsabs.harvard.edu/abs/2002A%26A...382...28S} {382, 28}

\bibitem[\protect\citeauthoryear{{Schaerer} \& {de Barros}}{{Schaerer} \& {de
  Barros}}{2009}]{Schaerer09}
{Schaerer} D.,  {de Barros} S.,  2009, \mn@doi [\aap]
  {10.1051/0004-6361/200911781}, \href
  {https://ui.adsabs.harvard.edu/abs/2009A&A...502..423S} {502, 423}

\bibitem[\protect\citeauthoryear{{Schaye} \& {Dalla Vecchia}}{{Schaye} \&
  {Dalla Vecchia}}{2008}]{Schaye08}
{Schaye} J.,  {Dalla Vecchia} C.,  2008, \mn@doi [\mnras]
  {10.1111/j.1365-2966.2007.12639.x}, \href
  {http://ads.nao.ac.jp/abs/2008MNRAS.383.1210S} {383, 1210}

\bibitem[\protect\citeauthoryear{{Schaye} et~al.,}{{Schaye}
  et~al.}{2010}]{Schaye10}
{Schaye} J.,  et~al., 2010, \mn@doi [\mnras]
  {10.1111/j.1365-2966.2009.16029.x}, \href
  {http://adsabs.harvard.edu/abs/2010MNRAS.402.1536S} {402, 1536}

\bibitem[\protect\citeauthoryear{{Schaye} et~al.,}{{Schaye}
  et~al.}{2015}]{Schaye15}
{Schaye} J.,  et~al., 2015, \mn@doi [\mnras] {10.1093/mnras/stu2058}, \href
  {http://adsabs.harvard.edu/abs/2015MNRAS.446..521S} {446, 521}

\bibitem[\protect\citeauthoryear{{Shang}, {Bryan}  \& {Haiman}}{{Shang}
  et~al.}{2010}]{Shang10}
{Shang} C.,  {Bryan} G.~L.,   {Haiman} Z.,  2010, \mn@doi [\mnras]
  {10.1111/j.1365-2966.2009.15960.x}, \href
  {https://ui.adsabs.harvard.edu/abs/2010MNRAS.402.1249S} {402, 1249}

\bibitem[\protect\citeauthoryear{{Shibuya}, {Kashikawa}, {Ota}, {Iye}, {Ouchi},
  {Furusawa}, {Shimasaku}  \& {Hattori}}{{Shibuya} et~al.}{2012}]{Shibuya12}
{Shibuya} T.,  {Kashikawa} N.,  {Ota} K.,  {Iye} M.,  {Ouchi} M.,  {Furusawa}
  H.,  {Shimasaku} K.,   {Hattori} T.,  2012, \mn@doi [\apj]
  {10.1088/0004-637X/752/2/114}, \href
  {http://ads.nao.ac.jp/abs/2012ApJ...752..114S} {752, 114}

\bibitem[\protect\citeauthoryear{{Smith}, {Wise}, {O'Shea}, {Norman}  \&
  {Khochfar}}{{Smith} et~al.}{2015}]{Smith15}
{Smith} B.~D.,  {Wise} J.~H.,  {O'Shea} B.~W.,  {Norman} M.~L.,   {Khochfar}
  S.,  2015, \mn@doi [\mnras] {10.1093/mnras/stv1509}, \href
  {https://ui.adsabs.harvard.edu/abs/2015MNRAS.452.2822S} {452, 2822}

\bibitem[\protect\citeauthoryear{{Song}, {Finkelstein}, {Livermore}, {Capak},
  {Dickinson}  \& {Fontana}}{{Song} et~al.}{2016}]{Song16b}
{Song} M.,  {Finkelstein} S.~L.,  {Livermore} R.~C.,  {Capak} P.~L.,
  {Dickinson} M.,   {Fontana} A.,  2016, \mn@doi [\apj]
  {10.3847/0004-637X/826/2/113}, \href
  {http://adsabs.harvard.edu/abs/2016ApJ...826..113S} {826, 113}

\bibitem[\protect\citeauthoryear{{Springel}}{{Springel}}{2005}]{Springel05e}
{Springel} V.,  2005, \mn@doi [MNRAS] {10.1111/j.1365-2966.2005.09655.x}, \href
  {http://adsabs.harvard.edu/abs/2005MNRAS.364.1105S} {364, 1105}

\bibitem[\protect\citeauthoryear{{Stacy} \& {Bromm}}{{Stacy} \&
  {Bromm}}{2014}]{Stacy14}
{Stacy} A.,  {Bromm} V.,  2014, \mn@doi [\apj] {10.1088/0004-637X/785/1/73},
  \href {http://adsabs.harvard.edu/abs/2014ApJ...785...73S} {785, 73}

\bibitem[\protect\citeauthoryear{{Sugimura}, {Matsumoto}, {Hosokawa}, {Hirano}
  \& {Omukai}}{{Sugimura} et~al.}{2020}]{Sugimura20}
{Sugimura} K.,  {Matsumoto} T.,  {Hosokawa} T.,  {Hirano} S.,   {Omukai} K.,
  2020, \mn@doi [\apjl] {10.3847/2041-8213/ab7d37}, \href
  {https://ui.adsabs.harvard.edu/abs/2020ApJ...892L..14S} {892, L14}

\bibitem[\protect\citeauthoryear{{Susa}, {Hasegawa}  \& {Tominaga}}{{Susa}
  et~al.}{2014}]{Susa14}
{Susa} H.,  {Hasegawa} K.,   {Tominaga} N.,  2014, \mn@doi [\apj]
  {10.1088/0004-637X/792/1/32}, \href
  {http://adsabs.harvard.edu/abs/2014ApJ...792...32S} {792, 32}

\bibitem[\protect\citeauthoryear{{Tamura} et~al.,}{{Tamura}
  et~al.}{2019}]{Tamura19}
{Tamura} Y.,  et~al., 2019, \mn@doi [\apj] {10.3847/1538-4357/ab0374}, \href
  {https://ui.adsabs.harvard.edu/abs/2019ApJ...874...27T} {874, 27}

\bibitem[\protect\citeauthoryear{{Topping}, {Stark}, {Endsley}, {Plat},
  {Whitler}, {Chen}  \& {Charlot}}{{Topping} et~al.}{2022}]{Topping22}
{Topping} M.~W.,  {Stark} D.~P.,  {Endsley} R.,  {Plat} A.,  {Whitler} L.,
  {Chen} Z.,   {Charlot} S.,  2022, arXiv e-prints, \href
  {https://ui.adsabs.harvard.edu/abs/2022arXiv220801610T} {p. arXiv:2208.01610}

\bibitem[\protect\citeauthoryear{{Tornatore}, {Ferrara}  \&
  {Schneider}}{{Tornatore} et~al.}{2007}]{Tornatore07}
{Tornatore} L.,  {Ferrara} A.,   {Schneider} R.,  2007, \mn@doi [\mnras]
  {10.1111/j.1365-2966.2007.12215.x}, \href
  {https://ui.adsabs.harvard.edu/abs/2007MNRAS.382..945T} {382, 945}

\bibitem[\protect\citeauthoryear{{Trebitsch}, {Blaizot}, {Rosdahl}, {Devriendt}
   \& {Slyz}}{{Trebitsch} et~al.}{2017}]{Trebitsch17}
{Trebitsch} M.,  {Blaizot} J.,  {Rosdahl} J.,  {Devriendt} J.,   {Slyz} A.,
  2017, \mn@doi [\mnras] {10.1093/mnras/stx1060}, \href
  {https://ui.adsabs.harvard.edu/abs/2017MNRAS.470..224T} {470, 224}

\bibitem[\protect\citeauthoryear{{Vanzella} et~al.,}{{Vanzella}
  et~al.}{2023}]{Vanzella23}
{Vanzella} E.,  et~al., 2023, \mn@doi [arXiv e-prints]
  {10.48550/arXiv.2305.14413}, \href
  {https://ui.adsabs.harvard.edu/abs/2023arXiv230514413V} {p. arXiv:2305.14413}

\bibitem[\protect\citeauthoryear{{Wise}, {Turk}, {Norman}  \& {Abel}}{{Wise}
  et~al.}{2012}]{Wise12a}
{Wise} J.~H.,  {Turk} M.~J.,  {Norman} M.~L.,   {Abel} T.,  2012, \mn@doi
  [\apj] {10.1088/0004-637X/745/1/50}, \href
  {http://ads.nao.ac.jp/abs/2012ApJ...745...50W} {745, 50}

\bibitem[\protect\citeauthoryear{{Wise}, {Demchenko}, {Halicek}, {Norman},
  {Turk}, {Abel}  \& {Smith}}{{Wise} et~al.}{2014}]{Wise14}
{Wise} J.~H.,  {Demchenko} V.~G.,  {Halicek} M.~T.,  {Norman} M.~L.,  {Turk}
  M.~J.,  {Abel} T.,   {Smith} B.~D.,  2014, \mn@doi [\mnras]
  {10.1093/mnras/stu979}, \href
  {http://adsabs.harvard.edu/abs/2014MNRAS.442.2560W} {442, 2560}

\bibitem[\protect\citeauthoryear{{Wise}, {Regan}, {O'Shea}, {Norman}, {Downes}
  \& {Xu}}{{Wise} et~al.}{2019}]{Wise19}
{Wise} J.~H.,  {Regan} J.~A.,  {O'Shea} B.~W.,  {Norman} M.~L.,  {Downes}
  T.~P.,   {Xu} H.,  2019, \mn@doi [\nat] {10.1038/s41586-019-0873-4}, \href
  {https://ui.adsabs.harvard.edu/abs/2019Natur.566...85W} {566, 85}

\bibitem[\protect\citeauthoryear{{Wolcott-Green}, {Haiman}  \&
  {Bryan}}{{Wolcott-Green} et~al.}{2011}]{Wolcott-Green11}
{Wolcott-Green} J.,  {Haiman} Z.,   {Bryan} G.~L.,  2011, \mn@doi [\mnras]
  {10.1111/j.1365-2966.2011.19538.x}, \href
  {https://ui.adsabs.harvard.edu/abs/2011MNRAS.418..838W} {418, 838}

\bibitem[\protect\citeauthoryear{{Wollenberg}, {Glover}, {Clark}  \&
  {Klessen}}{{Wollenberg} et~al.}{2020}]{Wollenberg20}
{Wollenberg} K. M.~J.,  {Glover} S. C.~O.,  {Clark} P.~C.,   {Klessen} R.~S.,
  2020, \mn@doi [\mnras] {10.1093/mnras/staa289}, \href
  {https://ui.adsabs.harvard.edu/abs/2020MNRAS.494.1871W} {494, 1871}

\bibitem[\protect\citeauthoryear{{Xu}, {Norman}, {O'Shea}  \& {Wise}}{{Xu}
  et~al.}{2016}]{Xu16}
{Xu} H.,  {Norman} M.~L.,  {O'Shea} B.~W.,   {Wise} J.~H.,  2016, \mn@doi
  [\apj] {10.3847/0004-637X/823/2/140}, \href
  {https://ui.adsabs.harvard.edu/abs/2016ApJ...823..140X} {823, 140}

\bibitem[\protect\citeauthoryear{{Yajima} \& {Khochfar}}{{Yajima} \&
  {Khochfar}}{2016}]{Yajima16a}
{Yajima} H.,  {Khochfar} S.,  2016, \mn@doi [\mnras] {10.1093/mnras/stw058},
  \href {http://adsabs.harvard.edu/abs/2016MNRAS.457.2423Y} {457, 2423}

\bibitem[\protect\citeauthoryear{{Yajima}, {Umemura}, {Mori}  \&
  {Nakamoto}}{{Yajima} et~al.}{2009}]{Yajima09}
{Yajima} H.,  {Umemura} M.,  {Mori} M.,   {Nakamoto} T.,  2009, \mn@doi [MNRAS]
  {10.1111/j.1365-2966.2009.15195.x}, \href
  {http://adsabs.harvard.edu/abs/2009MNRAS.398..715Y} {398, 715}

\bibitem[\protect\citeauthoryear{{Yajima}, {Choi}  \& {Nagamine}}{{Yajima}
  et~al.}{2011}]{Yajima11}
{Yajima} H.,  {Choi} J.-H.,   {Nagamine} K.,  2011, \mn@doi [\mnras]
  {10.1111/j.1365-2966.2010.17920.x}, \href
  {http://ads.nao.ac.jp/abs/2011MNRAS.412..411Y} {412, 411}

\bibitem[\protect\citeauthoryear{{Yajima}, {Li}, {Zhu}  \& {Abel}}{{Yajima}
  et~al.}{2012}]{Yajima12b}
{Yajima} H.,  {Li} Y.,  {Zhu} Q.,   {Abel} T.,  2012, \mn@doi [\mnras]
  {10.1111/j.1365-2966.2012.21228.x}, \href
  {http://ads.nao.ac.jp/abs/2012MNRAS.424..884Y} {424, 884}

\bibitem[\protect\citeauthoryear{{Yajima}, {Li}, {Zhu}, {Abel}, {Gronwall}  \&
  {Ciardullo}}{{Yajima} et~al.}{2014}]{Yajima14c}
{Yajima} H.,  {Li} Y.,  {Zhu} Q.,  {Abel} T.,  {Gronwall} C.,   {Ciardullo} R.,
   2014, \mn@doi [\mnras] {10.1093/mnras/stu299}, \href
  {http://adsabs.harvard.edu/abs/2014MNRAS.440..776Y} {440, 776}

\bibitem[\protect\citeauthoryear{{Yajima}, {Nagamine}, {Zhu}, {Khochfar}  \&
  {Dalla Vecchia}}{{Yajima} et~al.}{2017}]{Yajima17c}
{Yajima} H.,  {Nagamine} K.,  {Zhu} Q.,  {Khochfar} S.,   {Dalla Vecchia} C.,
  2017, \mn@doi [\apj] {10.3847/1538-4357/aa82b5}, \href
  {https://ui.adsabs.harvard.edu/abs/2017ApJ...846...30Y} {846, 30}

\bibitem[\protect\citeauthoryear{{Yajima}, {Sugimura}  \& {Hasegawa}}{{Yajima}
  et~al.}{2018}]{Yajima18}
{Yajima} H.,  {Sugimura} K.,   {Hasegawa} K.,  2018, \mn@doi [\mnras]
  {10.1093/mnras/sty997}, \href
  {https://ui.adsabs.harvard.edu/abs/2018MNRAS.477.5406Y} {477, 5406}

\bibitem[\protect\citeauthoryear{{Yajima} et~al.,}{{Yajima}
  et~al.}{2022}]{Yajima22}
{Yajima} H.,  et~al., 2022, \mn@doi [\mnras] {10.1093/mnras/stab3092}, \href
  {https://ui.adsabs.harvard.edu/abs/2022MNRAS.509.4037Y} {509, 4037}

\bibitem[\protect\citeauthoryear{{Yoon} et~al.,}{{Yoon} et~al.}{2022}]{Yoon22}
{Yoon} I.,  et~al., 2022, arXiv e-prints, \href
  {https://ui.adsabs.harvard.edu/abs/2022arXiv221008413Y} {p. arXiv:2210.08413}

\makeatother
\end{thebibliography}

\label{lastpage}

\end{document}